\def\mibS{} 
 
\documentclass[sigconf, nonacm, twocolumn]{acmart}

\usepackage{amsmath}
\usepackage{xfrac}
\usepackage{xcolor}
\usepackage{tikz}
\usetikzlibrary {shapes.multipart, shapes}
\usetikzlibrary{pgfplots.groupplots}
\usetikzlibrary{arrows.meta}
\usepackage[edges]{forest}
\usepackage{xurl}
\usepackage[ruled, vlined, linesnumbered, algo2e]{algorithm2e}
\usepackage{pgfplots}
\usepackage{graphicx}
\usepackage{adjustbox}
\usepackage{subcaption}
\usepackage{caption}
\usepackage{pgfplotstable}
\usepackage{footnote}
\makesavenoteenv{tabular}

\newcommand{\inreport}[2]{%
\ifdefined\mibS%
#1%
\else%
#2%
\fi}

\pgfplotscreateplotcyclelist{mylist}
{
    {red,     mark = *,         mark size = 0.80},
    {blue,    mark = square*,   mark size = 0.50},
    {brown,   mark = diamond*,  mark size = 0.80},
    {green,   mark = star,      mark size = 0.80},
    {black,   mark = triangle*,   mark size = 0.80},
}

\pgfplotscreateplotcyclelist{mylist2}
{
    {red},
    {blue},
    {brown},
}

\forestset{
  dtree/.style={
    for tree={
      font={\ttfamily \large},
      align=center,
      parent anchor=south,
      child anchor=north,
      s sep-=1mm,
    }
  },
  etype/.style n args={3}{
    edge path={
      \noexpand \path[\forestoption{edge},#3]
        (!u.parent anchor) -- (.child anchor)
        \forestoption{edge label};
      \noexpand \path[#1] (.child anchor) circle[radius=2pt];
      \noexpand \path[#2] (!u.parent anchor) circle[radius=2pt];
    },
  },
  dchild/.style={
    etype={fill=black}{draw=none}{black},
  },
}

\newcommand\vldbdoi{XX.XX/XXX.XX}
\newcommand\vldbpages{XXX-XXX}
\newcommand\vldbvolume{15}
\newcommand\vldbissue{11}
\newcommand\vldbyear{2022}
\newcommand\vldbauthors{\authors}
\newcommand\vldbtitle{\shorttitle} 
\newcommand\vldbavailabilityurl{URL_TO_YOUR_ARTIFACTS}

\inreport{ 
  \newcommand\vldbpagestyle{plain} 
}{ 
  \newcommand\vldbpagestyle{empty} 
}

\author{Qing Chen}
\affiliation{%
  \institution{University of Zurich}
}
\email{qing@ifi.uzh.ch}
\author{Oded Lachish}
\affiliation{%
  \institution{Birkbeck, University of London}
}
\email{o.lachish@bbk.ac.uk}
\author{Sven Helmer}
\affiliation{%
  \institution{University of Zurich}
}
\email{helmer@ifi.uzh.ch}
\author{Michael H. B{\"o}hlen}
\affiliation{%
  \institution{University of Zurich}
}
\email{boehlen@ifi.uzh.ch}

\usepackage{xcolor,colortbl}
\usepackage{mdframed}

\definecolor{Gray}{gray}{0.98}

\newmdenv[
  topline=false,
  bottomline=false,
  rightline=false,
  leftline=true,
  skipabove=\topsep,
  skipbelow=\topsep,
  backgroundcolor=gray!3
]{siderules}

\begin{document}

\edef\oldindent{\the\parindent}
\edef\oldparskip{\the\parskip}
\setlength{\parindent}{0em}
\setlength{\parskip}{2ex}
\newcommand\shorttitle{} 
\renewcommand\shortauthors{} 
\pagestyle{plain}


\setlength{\parindent}{\oldindent}%
\setlength{\parskip}{\oldparskip}


\title{Dynamic Spanning Trees for Connectivity Queries on
  Fully-dynamic Undirected Graphs\inreport{ (Extended Version)}{}}

\begin{abstract}
  Answering connectivity queries is fundamental to fully dynamic
  graphs where edges and vertices are inserted and deleted frequently.
  Existing work proposes data structures and algorithms with worst
  case guarantees.  We propose a new data structure, the \emph{dynamic
    tree} (D-tree), together with algorithms to construct and maintain
  it.  The D-tree is the first data structure that scales to
  fully dynamic graphs with millions of vertices and edges and, on
  average, answers connectivity queries much faster than data
  structures with worst case guarantees.
\end{abstract}

\maketitle

\pagestyle{\vldbpagestyle}
\begingroup\small\noindent\raggedright\textbf{PVLDB Reference Format:}\\
\vldbauthors. \vldbtitle. PVLDB, \vldbvolume(\vldbissue): \vldbpages, \vldbyear.\\
\href{https://doi.org/\vldbdoi}{doi:\vldbdoi} \endgroup \begingroup
\renewcommand\thefootnote{}\footnote{\noindent
  This work is licensed under the Creative Commons BY-NC-ND 4.0 International License.
  Visit \url{https://creativecommons.org/licenses/by-nc-nd/4.0/} to view a copy of this
  license. For any use beyond those covered by this license, obtain permission by
  emailing \href{mailto:info@vldb.org}{info@vldb.org}. Copyright is held by the
  owner/author(s). Publication rights licensed to the VLDB Endowment. \\
  \raggedright Proceedings of the VLDB Endowment, Vol. \vldbvolume,
  No. \vldbissue\ %
  ISSN 2150-8097. \\
  \href{https://doi.org/\vldbdoi}{doi:\vldbdoi} \\
}\addtocounter{footnote}{-1}\endgroup

\ifdefempty{\vldbavailabilityurl}{}{ \vspace{.3cm}
  \begingroup\small\noindent\raggedright\textbf{PVLDB Artifact Availability:}\\
  The source code, data, and/or other artifacts have been made
  available at 
  
  \url{https://github.com/qingchen3/D-tree}.  
  \endgroup }

\setcounter{figure}{0}

\section{Introduction}
\label{sec:intro}

The efficient processing of large graphs is becoming ever more
important (see Hegeman and Iosup \cite{Hegeman18}, Sahu et al.\
\cite{vldb_largegraphs}, and Sakr et al.\ \cite{Sakr21} for recent
studies and surveys).  A fundamental problem is the connectivity
problem, which checks if there is a connection between two nodes in a
graph.  Answering connectivity queries plays a crucial role in
application areas such as communication and transport networks,
checking their reliability, as well as social networks, investigating
the connections between users and the groups they belong to.
However, it does not stop there: since dynamic connectivity is such a
fundamental problem, we find applications in areas as diverse as
computational geometry~\cite{doraiswamy2009efficient},
chemistry~\cite{eyal2005improved},
and biology~\cite{henzinger1999constructing}.

Computing the connectivity between two nodes using search strategies
like breadth-first search (BFS) and depth-first search (DFS) with a
linear run-time is prohibitively expensive for large graphs with
millions of vertices and edges.  For static graphs, the connected
components can be precomputed and the results stored in an auxiliary
data structure, allowing the efficient processing of queries.
Updating the auxiliary data structures in the fully dynamic case with
frequent graph edge insertions and deletions is challenging, though.
For instance, updating the well-known two-hop labeling
\cite{TOL,bramandia2009incremental,Labeling_2hop, DBL} is expensive,
since BFS or DFS must be run on the graphs.  Similarly,
tree-based approaches \cite{Fred83,henzinger99,Holm2001,Thorup2000,
  wulff2013faster, huang2017fully} have focused on worst-case runtime
guarantees and incur high update costs for large graphs.  They rely on
multiple complex auxiliary data structures, have often not been
implemented and evaluated empirically \cite{David1997experiment,
  Zaroliagis02}, and sacrifice average case performance to get an
upper bound for the worst-case complexity.  In our work, we focus on
fully dynamic large real-world graphs with the goal of developing a
connectivity algorithm with a good average case performance for
queries and updates.

First, we define what optimizing the average case complexity for
connectivity queries over the spanning forest (i.e., sets of spanning trees)
of a graph means: the costs are minimized if $S_d$, the sum of distances
between the root nodes and all the other nodes in the spanning trees, is
minimized. Since maintaining a minimal $S_d$ in spanning trees in a fully
dynamic setting is too expensive, we propose effective and practical
heuristics to keep the value of $S_d$ of the spanning trees low. Our
approach has a much better average
runtime than solutions with a guaranteed worst case complexity for a broad
range of real-word graphs (we demonstrate this empirically). 

The most time-critical part is the search for a replacement edge when 
deleting an edge in a
spanning tree.  We prove that the cost for finding a replacement edge
for an edge $e$ is proportional to the cut number of $e$, i.e., the
number of nodes in the smaller tree after removing $e$ (deleting an edge
splits a tree into two). Moreover, we
prove that the average cost of finding a replacement edge is optimal
for spanning trees that minimize $S_c$, the sum of the cut numbers for
all possible edges in the spanning tree.
We show that $S_d$ and $S_c$ are directly related to each other, i.e.,
  optimizing one also optimizes the other.


Our main technical contribution can be summarized as
follows:
\begin{itemize}
\item We formally define the problem of evaluating connectivity
  queries in fully dynamic graphs with an optimal average-case
  complexity.
\item We introduce $S_d$ and $S_c$. $S_d$ is the sum of distances
  between roots and all other nodes; we show that the average cost
  of connectivity queries is optimal for spanning forests minimizing
  $S_d$. $S_c$ is the sum of cut numbers of all edges; we show that
  the average costs for finding replacement edges is optimal if
  spanning trees minimize $S_c$.
\item We prove that $S_d = S_c$ for spanning trees in which the root
  is a centroid, i.e., a node that minimizes the sum of the distances
  to all other nodes, allowing us to optimize the average-case costs.
\item We propose a novel k-ary tree, called dynamic tree (D-tree), to
  represent the connected components of a graph.  We define D-trees
  and provide efficient, heuristics-based algorithms to answer 
  connectivity queries and
  maintain D-trees when inserting and deleting edges.
\item We embed the graph in a set of D-trees that also maintain edges
  not part of the spanning forest and the size of each subtree. This
  information helps us to keep the average runtimes of operations low.
\item We conduct extensive experiments to compare D-trees with
  existing approaches over ten real-world datasets.  The experiments
  confirm the efficiency of our approach and its superior average-case
  runtime.
\end{itemize}

\section{Related Work}
\label{sec:related}

The first efficient connectivity algorithms focused on updating
spanning trees in incremental~\cite{Tarjan75} and
decremental~\cite{Shiloach81} dynamic graphs, i.e., graphs only
allowing insertions or deletions, respectively.  The earliest
algorithms for updating minimum spanning trees in fully dynamic
undirected (weighted) graphs were developed by Spira and Pan
\cite{Spira75}, Chin and Houck \cite{Chin78}, and Frederickson
\cite{Fred83}.  The algorithm by Spira and Pan has a complexity of
$O(n)$ for insertions and $O(n^3)$ for deletions, with $n$ being the
number of vertices.  Chin and Houck improve the complexity for
deletions to $O(n^2)$. Frederickson brings the complexity of
insertions and deletions down to $O(\sqrt{m})$, with $m$ being the
number of edges.  Using a technique called sparsification, Eppstein et
al.\ improve the complexity to $O(\sqrt{n})$ per update
operation~\cite{Eppstein92,Eppstein97}, but without providing an
implementation.

Henzinger and King represent spanning trees via Euler tours
\cite{tarjan1984finding}, resulting in elegant merging and splitting
of spanning trees
\cite{henzinger1995randomized,henzinger97,henzinger99,henzinger01}.
Storing, searching, and maintaining Euler tours efficiently is not
trivial, though.  Henzinger and King proposed the Euler Tour Tree
(ET-tree) \cite{henzinger1995randomized, henzinger99} that maps Euler
tours to balanced binary trees \cite{David1997experiment,
  seidel1996randomized} and requires several auxiliary data structures
\cite{henzinger1995randomized, henzinger99} to keep track of
information for Euler tours.

The work by Henzinger and King~\cite{henzinger1995randomized,
  henzinger99} sparked a whole line of research based on hierarchical
forests for dynamic connectivity. We
divide the algorithms into two groups: those that minimize the
worst-case costs and those that optimize the amortized costs. We first
look at worst-case costs for update operations. Interestingly enough,
for sparse graphs, the algorithm by Frederickson~\cite{Fred83} (and
the improvement by Eppstein~\cite{Eppstein97}) is still
competitive. Kapron et al.~\cite{Kapron13} proposed an algorithm with
complexity $O(\log^5 n)$, but it turned out that it can produce false
negatives. In 2016, Kejlberg{-}Rasmussen et al.~\cite{Kejlberg16}
improved the complexity to
$O(\sqrt{\sfrac{n (\log \log n)^2}{\log n}})$. Henzinger and King were
the first to look at amortized costs and achieve polynomial
logarithmic amortized complexity. Holm at al.~\cite{Holm2001} improved
the bound by adding invariants to the hierarchical forests.
Orthogonal data structures, such as local trees, lazy local trees,
bitmaps, and a system of shortcuts~\cite{wulff2013faster, Thorup2000,
  huang2017fully}, are introduced to improve the amortized complexity.
The combination of these complicated data structures makes it
difficult to implement (and evaluate) these algorithms.  In fact, only
Henzinger-King's algorithm $HK$
\cite{henzinger1995randomized,henzinger99} was fully implemented and
evaluated \cite{Zaroliagis02, David1997experiment,Raj2001experiment}
and is therefore our main contender.

Most existing work on labeling schemes \cite{TOL,
  bramandia2009incremental, IP_label, TF_label, 3hop} requires that input
graphs are directed and/or DAGs, and consequently are generally not
applicable to undirected graphs. A recent data structure for labeling,
called DBL~\cite{DBL}, works for undirected
graphs. However, DBL only supports insertions on graphs, and constructing
DBL is expensive since it needs to run BFS on connected components.

\section{Preliminaries}
\label{sec:prelim}


We consider \emph{undirected unweighted simple graphs} $G(V,E)$
defined by a set of vertices $V$ and a set $E$ of edges
\cite{gibbons1985algorithmic, west2001introduction}.  A graph is
\emph{simple} iff there is at most one edge $(u,v) \in E$ that
connects a pair of vertices $u,v \in V$. We measure the \emph{size} of
a graph in the number of vertices it contains, which we denote by
$|V|$.  Given a graph $G(V,E)$, a \emph{path} $P$ is a sequence of
distinct vertices $(v_1, v_2, \dots, v_n)$, $v_i \in V$, such that
each pair of adjacent vertices in $P$, $v_i$ and $v_{i+1}$, are
connected via an edge $(v_i, v_{i+1}) \in E$.  The \emph{length} $|P|$
of a path $P$ is defined by the number of edges in the path, i.e., for
$P = (v_1, v_2, \dots, v_n)$, $|P| = n-1$.  If there is an additional
edge between $v_n$ and $v_1$, then the sequence
$(v_1, v_2, \dots, v_n)$ forms a \emph{cycle}. The \emph{diameter} of
a graph is the length of the longest shortest path between two
vertices in the graph.  A \emph{connected component} $C(V',E')$ is a
maximal subgraph of a graph $G(V,E)$, with
$V' \subseteq V, E' \subseteq E$, in which all pairs of nodes are
connected via a path.

\begin{figure}[htb]\centering
  \begin{subfigure}[b]{0.4\columnwidth}\centering
    \begin{tikzpicture}
      \tikzstyle{every node}=[fill=black,circle,inner sep=0pt,minimum size=2.5pt,]
      \node [label=above:$\mathsf{n}_{1}$] (1) at (-4.0, 0.0) {};
      \node [label=left:$\mathsf{n}_{2}$] (2) at (-5.0, -0.5) {};
      \node [label=left:$\mathsf{n}_{3}$] (3) at (-4.0, -0.5) {};
      \node [label=right:$\mathsf{n}_{4}$] (4) at (-3.0, -0.5) {};
      \node [label=below:$\mathsf{n}_{5}$] (5) at (-4.5, -1.0) {};
      \node [label=below:$\mathsf{n}_{6}$] (6) at (-3.5, -1.0) {};
      
      \draw[-] (1) to (2);
      \draw[-] (1) to (3);
      \draw[-] (1) to (4);
      \draw[-] (3) to (5);
      \draw[-] (2) to (5);
      \draw[-] (3) to (6);
      \draw[-] (4) to (6);
      \draw[-] (4) to (5);
    \end{tikzpicture}
    \caption{Component $C_1$}
  \end{subfigure}
  \begin{subfigure}[b]{0.4\columnwidth}\centering
    \begin{tikzpicture}
      \tikzstyle{every node}=[fill=black,circle,inner sep=0pt,minimum size=2.5pt,]
      \node [label=left:$\mathsf{n}_{10}$] (10) at (-2.0, -1.3) {};
      \node [label=right:$\mathsf{n}_{11}$] (11) at (-1.0, -1.3) {};
      \node [label=above:$\mathsf{n}_{9}$] (9) at (-1.4, -0.9) {};
      \node [label=below:$\mathsf{n}_{12}$] (12) at (-1.4, -1.7) {};

      \draw[-] (9) to (10);
      \draw[-] (9) to (11);
      \draw[-] (10) to (11);
      \draw[-] (10) to (12);
    \end{tikzpicture}
    \caption{Component $C_2$}
  \end{subfigure}
  \begin{tikzpicture}[remember picture,overlay]
    \node at (-3.0,2.4) {Graph $G_1$};
    \draw [dotted] plot [smooth cycle] coordinates {
      (-2,.5)(-3.5,1.2)(-4.5,.5)(-7,.8)(-6.5,1.9)(-5,2.5) (-3,2.1)(-1,2.4)(-.5,1)
    };
  \end{tikzpicture}
  \caption{$G_1 = \{ C_1, C_2\}$ with components $C_1$ and
    $C_2$} \label{fig:example_graph_G1}
\end{figure}
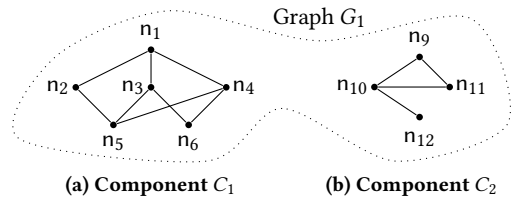

\begin{example}\label{example:comp}
  Figure~\ref{fig:example_graph_G1} shows a graph $G_1$ with two connected
  components $C_1$ and $C_2$. 
\end{example}

A \emph{tree} is an undirected graph in which any pair of vertices is
connected by \emph{exactly} one path. Thus, the vertices in a tree are
all connected and the tree does not contain cycles. In a
\emph{forest}, any two vertices are connected by \emph{at most} one
path, which means that its connected components consist of trees.  In
a \emph{rooted tree}, we designate one vertex as the root $r$ of the
tree. By definition, $r$ has depth 0. The \emph{depth} of any other
vertex $v$ is determined by its \emph{(tree) distance} $d_T$ to $r$,
i.e., $d_T(r,v)$ is equal to the length of the path from the root to
the vertex. The \emph{height} of a tree is equal to the depth of the
  leaf node with the maximum depth. Given a rooted
tree with root $r$, the \emph{ancestors}, $anc(v)$, of a node
$v \not= r$ ($r$ does not have any ancestors) are all the nodes on the
path from $v$ to $r$ except $v$.  The \emph{parent} of $v$ is the node
$u$ on this path with $depth(u) + 1 = depth(v)$.  The \emph{children}
of $v$ are the nodes that have $v$ as a parent.  The
\emph{descendants}, $desc(v)$, of $v$ are all nodes $u \not= v$ for
which $v$ appears in the path from $u$ to $r$.  The \emph{subtree}
rooted at $v$ consists of $v$ and all its descendants.  The
\emph{size} of this subtree, denoted by $size(v)$, is measured in the
number of nodes it includes.  Given a connected component $C(V',E')$,
a \emph{spanning tree} $T = (V',E_T)$, with $E_T \subseteq E'$, is a
rooted tree containing all vertices of $C$. We use a \emph{spanning
  forest}, consisting of a spanning tree for each component, for
graphs with more than one component.

\begin{figure}[htb]\centering
  \begin{subfigure}[b]{0.4\columnwidth}\centering
    \begin{tikzpicture}
      \tikzstyle{every node}=[fill=black,circle,inner sep=0pt,minimum size=2.5pt,]
      \node [fill=red, label=above:$\mathsf{n}_{1}$] (1) at (-4.0, 0.0) {};
      \node [label=left:$\mathsf{n}_{2}$] (2) at (-5.0, -0.5) {};
      \node [label=left:$\mathsf{n}_{3}$] (3) at (-4.0, -0.5) {};
      \node [label=right:$\mathsf{n}_{4}$] (4) at (-3.0, -0.5) {};
      \node [label=below:$\mathsf{n}_{5}$] (5) at (-4.5, -1.0) {};
      \node [label=below:$\mathsf{n}_{6}$] (6) at (-3.5, -1.0) {};
      
      \draw[-] (1) to (2);
      \draw[-] (1) to (3);
      \draw[-] (1) to (4);
      \draw[-] (3) to (5);
      \draw[-] (4) to (6);
    \end{tikzpicture}
    \caption{Spanning Tree $T_1$ for $C_1$}
  \end{subfigure}
  \quad
  \begin{subfigure}[b]{0.4\columnwidth}\centering
    \begin{tikzpicture}    
      \tikzstyle{every node}=[fill=black,circle,inner sep=0pt,minimum size=2.5pt,]
      \node [label=left:$\mathsf{n}_{10}$] (10) at (-2.0, -1.3) {};
      \node [label=right:$\mathsf{n}_{11}$] (11) at (-1.0, -1.3) {};
      \node [fill=red, label=above:$\mathsf{n}_{9}$] (9) at (-1.4, -0.9) {};
      \node [label=below:$\mathsf{n}_{12}$] (12) at (-1.4, -1.7) {};

      \draw[-] (9) to (10);
      \draw[-] (9) to (11);
      \draw[-] (10) to (12);
    \end{tikzpicture}
    \caption{Spanning Tree $T_2$ for $C_2$}
  \end{subfigure}
  \begin{tikzpicture}[remember picture,overlay]
    \node at (-3.0,2.4) {Forest $F_1$};
    \draw [dotted] plot [smooth cycle] coordinates {
      (-2,.5)(-3.5,1.2)(-4.8,.5)(-6.5,.6)(-7.2,1.8)(-6.5,2.3)(-5,2.5)(-3,2.1)(-1,2.4)(-.5,1)
    };
  \end{tikzpicture}
  \caption{Spanning forest $F_1 = \{ T_1, T_2 \}$ for $G_1$ with
    spanning trees $T_1$ and $T_2$ for components $C_1$ and $C_2$. The
    roots of the spanning trees are colored red.}
  \label{fig:graph_spanning_forest}
\end{figure}
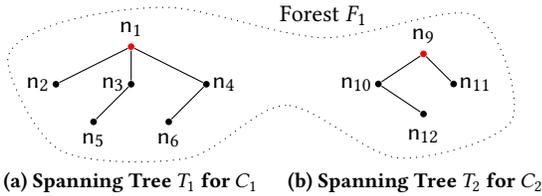

\begin{example}\label{example:comp}
  Figure~\ref{fig:graph_spanning_forest} depicts spanning forest $F_1$
  for graph $G_1$ from Figure~\ref{fig:example_graph_G1}. $F_1$ is
  made up of spanning trees $T_1$ and $T_2$ for components $C_1$ and
  $C_2$, respectively. The path from $\mathsf{n_5}$ to $\mathsf{n_1}$
  is ($\mathsf{n_5}$, $\mathsf{n_3}$, $\mathsf{n_1}$);
  $anc(\mathsf{n_5})$ = $\{\mathsf{n_1}, \mathsf{n_3}\}$;
  $desc(\mathsf{n_3})$ = $\{\mathsf{n_5}\}$; $depth(\mathsf{n_3})$ = 1
  and $depth(\mathsf{n_5})$ = 2.  The subtree rooted at $\mathsf{n_3}$
  consists of $\mathsf{n_3}$ and its descendant $\mathsf{n_5}$, and
  the size of this subtree is 2.
\end{example}

\begin{definition}[Vertex deviation and centroid]
  \label{def:centroid}
  Given a tree $T = (V',E_T)$, the \emph{vertex deviation} $m(v)$ of a
  vertex $v \in V'$ is the average distance of $v$ to all other nodes:
  $m(v) = \sfrac{1}{|V'|} \sum_{u \in V'} d_T(v,u)$. A \emph{centroid}
  (or \emph{vertex median}) of $T$ is a vertex with minimal $m(v)$ for
  $T$~\cite{Jordan1869,Zelinka68}.
\end{definition}

A tree with an even number of vertices can have two centroids. In this
case, the two centroids are adjacent to each other~\cite{Zelinka68}.

\begin{example}
  The centroid of $T_1$ in Figure \ref{fig:graph_spanning_forest}(a)
  is $\mathsf{n_1}$ since the vertex deviation
  $m(\mathsf{n_1}) = \sfrac{(1 + 1 + 1 + 2 + 2)}{6} = \sfrac{7}{6}$,
  which is minimal for this tree.
\end{example}

\section{Problem Definition}
\label{sec:problemdef}

We now formally define \emph{connectivity queries} on
graphs and formulate the challenges posed by dynamic graphs.

\begin{definition}[Connectivity query]
  \label{def:conn}
  Given a graph $G(V,E)$ and two vertices $u,v \in V$, the
  \emph{connectivity query} $conn(u, v)$ returns True if there exists
  a path between $u$ and $v$ in $G$, and False otherwise.
\end{definition}

\begin{example}\label{example:conn_graph}
  Consider graph $G_1$ in Figure \ref{fig:example_graph_G1}. The
  connectivity query $conn(\mathsf{n_2}, \mathsf{n_6})$ returns True,
  as $\mathsf{n_2}$ and $\mathsf{n_6}$ are connected via
  $\mathsf{n_1}$ and $\mathsf{n_4}$ (and also via $\mathsf{n_5}$ and
  $\mathsf{n_3}$).  The connectivity query
  $conn(\mathsf{n_6}, \mathsf{n_9})$ returns False, because
  $\mathsf{n_6}$ and $\mathsf{n_9}$ are located in different
  components.
\end{example}

A naive approach for checking connectivity is
to run a search algorithm, such as breadth-first search (BFS) or
depth-first search (DFS), from one of the two vertices and test if
the search finds the other node, which is prohibitively expensive for large
graphs (it has complexity $O(|V| + |E|)$).
For static graphs, we can determine all connected components
of a graph, using BFS or DFS (see, e.g., ~\cite{Hop73}), and then
label the nodes with the ID of the component they belong to. Given two
nodes, we then directly decide in constant time whether they are
connected.  Evaluating connectivity queries on dynamic graphs
is a much more challenging scenario. We first formally
define dynamic graphs:

\begin{definition}[Fully dynamic graph]
  \label{def:dynamicgraph}
  In a \emph{fully dynamic graph} $G_d(V,E)$, edges are inserted and
  deleted one at a time.  We apply a sequence of update operations to
  a graph, $((t_1, o_1), (t_2, o_2),$ $(t_3, o_3), \dots)$, where
  $t_i$ is a timestamp and $o_i$ is either an insertion
  ($E_{t+1} = E_t \cup (v_i, v_j)$) or a deletion
  ($E_{t+1} = E_t \setminus (v_i, v_j)$) of an edge.
\end{definition}

Since we only deal with dynamic graphs from here on, we drop the
subscript $d$ and refer to dynamic graphs as $G(V,E)$. Our
implementation allows the insertion and deletion of isolated, i.e.,
unconnected vertices.  However, since spanning trees consisting
  of a single node are trivial to handle, we restrict our description
  to edge insertions and deletions.

As we will see later, in the worst case the performance of 
deletion operations is especially problematic.
We argue that these cases rarely occur in real-world graphs and that
it is more important to consider the average-case complexity.

Before going into the implementation details of our approach, which is based
on spanning trees, we explicitly define the problem we are solving in
Definition~\ref{def:problemdef} and then
investigate important aspects of applying spanning trees to evaluate
connectivity queries in fully dynamic graphs and show how we exploit these
properties in the following section.

\begin{definition}[Problem definition]
  \label{def:problemdef}
  Find a data structure that in fully dynamic graphs, on average, 
  allows us to (a) answer connectivity queries and (b) maintain the data 
  structure efficiently. 
\end{definition}

\section{Leveraging Spanning Trees}
\label{sec:spanningtree}

We first define the problem of evaluating connectivity queries with an
optimal average-case complexity. Next, we introduce $S_d$, which
optimizes average costs for connectivity queries, and $S_c$, which
optimizes average costs for searching for replacement edges.
Finally, we formally establish the relationship between $S_c$ and
$S_d$.  \inreport{All proofs for the theorems and lemmas in this
  section are shown in the appendix.}  {All proofs for the theorems
  and lemmas in this section are included in the technical
  report~\cite{technicalreport}.}

\subsection{Evaluating Queries}
\label{sec:evaluating_queries}

We use a spanning forest to answer connectivity queries $conn(u, v)$
by traversing the
paths from $u$ and $v$ to the respective roots $r_u$ and $r_v$ of
their spanning trees. If we end up at the same root, then $u$ and $v$
are located in the same component and are connected.  If we reach
different roots, they are not connected.  The costs for evaluating a
connectivity query $conn(u, v)$ via spanning trees is equal to the sum
of distances of $u$ and $v$ to their roots: $d_T(r_u,u) + d_T(r_v,v)$.

\begin{definition}[Sum of distances between root and its descendants]
  \label{def:sumDist}
  Given a (spanning) tree $T = (V', E_T)$ with root $r$, the sum of
  distances between $r$ and its descendants, $S_d$ is defined as
  follows:
  \begin{eqnarray}
    S_d(T) = \sum_{x \in V'} d_T(r, x).  \label{eq:s_d}
  \end{eqnarray}
\end{definition}

Before analyzing the average-case costs, we give a formal definition
of these costs:

\begin{definition}[Average-case complexity]
  \label{def:avgcase}
  Let $I$ be the set of all possible inputs for an algorithm $A$ and
  let $t(i)$, $i \in I$, be the cost of running $A$ on input $i$. The
  probability that input $i$ occurs is defined by $p(i)$. The
  \emph{average cost} of running $A$ is the expected value of the
  running times: $E(t) = \sum_{i \in I} t(i) p(i)$. If the
  probabilities $p(i)$ are not available, often a uniform distribution
  is assumed: $E(t) = \frac{1}{|I|} \sum_{i \in I} t(i)$.
\end{definition}

A workload-aware analysis utilizing the probability distribution of
the inputs is beyond the scope of this paper. 
In the following, we assume a uniform distribution of
the inputs.  We illustrate with an example what average-case versus
worst-case costs mean for connectivity queries.

\begin{figure}[htb]\centering
  \begin{subfigure}[b]{0.45\columnwidth}\centering
    \begin{tikzpicture}
      \tikzstyle{every node}=[fill=black,circle,inner sep=0pt,minimum size=2.5pt,]
      \node [fill=red, label=above:$\mathsf{n}_{1}$] (1) at (3.0, 0.0) {};
      \node [label=below:$\mathsf{n}_{2}$] (2) at (1.75, -0.5) {}; 
      \node [label=below:$\mathsf{n}_{3}$] (3) at (2.25, -0.5) {};
      \node [label=below:$...$, fill=white] (4) at (2.85, -0.5) {};
      \node [label=right:$\mathsf{n}_{15}$] (15) at (3.35, -0.5) {};

      \node [label=right:$\mathsf{n}_{16}$] (16) at (4.5, -0.5) {};
      \node [label=left:$\mathsf{n}_{17}$] (17) at (4.2, -1.0) {};
      \node [label=right:$\mathsf{n}_{18}$] (18) at (4.8, -1.0) {}; 
      \node [label=left:$\mathsf{n}_{19}$] (19) at (3.9, -1.5) {};
      \node [label=right:$\mathsf{n}_{20}$] (20) at (5.1, -1.5) {}; 
      
      \draw[-] (1) to (2);
      \draw[-] (1) to (3);
      \draw[-] (1) to (3.00, -0.5);
      \draw[-] (1) to (2.75, -0.5);
      \draw[-] (1) to (16);
      \draw[-] (1) to (15);
      \draw[-] (16) to (17);
      \draw[-] (16) to (18);
      \draw[-] (17) to (19);
      \draw[-] (18) to (20);

    \end{tikzpicture}

    \caption{Structure of $T_1$, $S_d$ = 25.}
  \end{subfigure}
  \quad
  \begin{subfigure}[b]{0.45\columnwidth}\centering
    \begin{tikzpicture}
      \tikzstyle{every node}=[fill=black,circle,inner sep=0pt,minimum size=2.5pt,]
      \node [fill=red, label=above:$\mathsf{n}_{16}$] (16) at (3.5, 0.0) {};
      \node [label=above:$\mathsf{n}_{1}$] (1) at (2.75, -0.5) {};

      \node [label=below:$\mathsf{n}_{2}$] (2) at (2.0, -1.0) {};
      \node [label=below:$\mathsf{n}_{3}$] (3) at (2.5, -1.0) {};
      \node [label=below:$\mathsf{...}$, fill=white] (4) at (3.0, -1.0) {};
      \node [label=right:$\mathsf{n}_{15}$] (5) at (3.5, -1.0) {};
      
      \node [label=left:$\mathsf{n}_{17}$] (17) at (4.0, -0.5) {};
      \node [label=right:$\mathsf{n}_{18}$] (18) at (4.5, -0.5) {};

      \node [label=below:$\mathsf{n}_{19}$] (19) at (4.5, -1.0) {}; 
      \node [label=below:$\mathsf{n}_{20}$] (20) at (5.2, -1.0) {};  
      
      \draw[-] (1) to (2);
      \draw[-] (1) to (3);
      \draw[-] (1) to (3.05, -1.0);
      \draw[-] (1) to (2.85, -1.0);
      \draw[-] (1) to (5);
      \draw[-] (1) to (16);
      \draw[-] (2) to (1);
      \draw[-] (16) to (17);
      \draw[-] (16) to (18);
      \draw[-] (17) to (19);
      \draw[-] (18) to (20);

    \end{tikzpicture}
    \caption{Balanced trees $T'_1$, $S_d$ = 35.}
  \end{subfigure}
  \caption{Unbalanced versus balanced spanning trees}
  \label{fig:avgworsttree}
\end{figure}
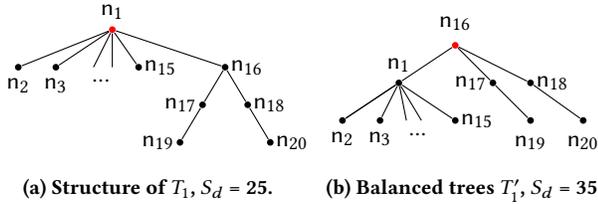

\begin{example}\label{example:avgworstcase}
  Consider the spanning tree $T_1$ in
  Figure~\ref{fig:avgworsttree}(a).  Then the worst case for
  evaluating a connectivity query occurs if we select
  $T_1.n_{19}$ and $T_1.n_{20}$ as parameters, leading to a cost of
  $3+3=6$. Assuming a uniform distribution of inputs for connectivity
  queries on $T_1$, we get
  $\sfrac{2 * S_d(T_1)}{|T_1|} = \sfrac{(2 * 25)}{20} = 2.5$ for the
  average costs. If we balance the tree by rerooting it, we get $T'_1$
  as shown in Figure~\ref{fig:avgworsttree}(b).  For $T'_1$ the costs
  are $4$ in the worst case and $3.5$ in the average case.
\end{example}

In Example~\ref{example:avgworstcase}, by balancing the spanning trees (and
optimizing the worst case), we actually worsen the average costs. Looking at $T_1$ in
Figure~\ref{fig:avgworsttree}(a), we can see that the paths from $n_1$ to 
$n_{19}$ and from $n_1$ to $n_{20}$ are outliers, all the other nodes are 
very close to $n_1$. In essence,
balancing the tree punishes the performance of all other queries not involving
these outliers. For this reason, other (tree-like) data structures, such as
tries~\cite{Szpan90} and multilevel extendible hashing
schemes~\cite{Helmer03}, do not strive for balance, but allow the
outlier parts to grow deeper than the rest of the tree.

We now investigate what spanning trees have to look like to guarantee
minimum average costs.

\begin{theorem} \label{theorem:average_cost} The average costs of
  evaluating connectivity queries with spanning trees is optimal if
  the trees in the spanning forest minimize $S_d$.
\end{theorem}

\inreport{
  \begin{proof}
  Shown in Appendix \ref{average_cost_proof}.
  \end{proof}
}

Generally, a high fanout leads to shallow trees (B-trees are a
classical example), which in turn decreases the distances between the
root and other nodes. When it comes to spanning trees, using
breadth-first-search (BFS) trees provides excellent fanout, minimizing
$S_d$ for a given root.

\begin{definition}[Breadth-first-search tree (BFS-tree)]
  For a connected component $C = (V', E')$ (or a connected graph), a
  BFS-tree is a spanning tree constructed by a breadth first
  search, which traverses the component level by level, starting from
  the root node $r$ of the BFS-tree, then visiting all the nodes at a
  distance of one, at a distance of two, and so on.
\end{definition}

\inreport{
  \begin{example}
    Shown in Appendix \ref{exam_spanning_trees}
  \end{example}
}

\begin{lemma} \label{lemma:bfs_tree} %
  In a BFS-tree with root $r$ the sum of distances $S_d$ between $r$
  and all other nodes is minimal.
\end{lemma}

\inreport{
  \begin{proof}
    Shown in Appendix \ref{bfs_tree_proof}
  \end{proof}
}

So, we could compute the optimal BFS-tree for each component, i.e., if
$P = \{\mbox{BFS-tree with root } v|v \in V'\}$ is the the set of all
BFS-trees with different roots for component $C = (V', E')$, we select
the tree with $S_d = \min_{T \in P} S_d(T)$ .  This
optimizes the average cost of running connectivity queries via
spanning trees.  For fully dynamic graphs, it is too expensive to
update these spanning trees while preserving 
them to be optimal BFS-trees. Instead, we switch to efficient
  heuristics, e.g., by picking a root that is a centroid.

\subsection{Updating Spanning Trees}
\label{sec:updating}

We distinguish two different types of edges in a connected component:
those that belong to the current spanning tree representing the
component, which we call \emph{tree edges}, and those that do not,
which we call \emph{non-tree edges}.

\begin{definition}[Tree and non-tree edges]
  \label{def:treeedges}
  Consider a connected component $C(V',E')$ and a spanning tree
  $T = (V', E_T)$ for $C$.  An edge $(u, v) \in E'$ is a \emph{tree
    edge} for $C$ if $(u, v) \in E_T$, and a \emph{non-tree edge} for
  $C$ if $(u, v) \in E' \setminus E_T$.
\end{definition}

\begin{example}
  Consider component $C_1 = (V_1, E_1)$
  in Figure~\ref{fig:example_graph_G1}(a) and
  spanning tree $T_1$ for $C_1$ in Figure~\ref{fig:graph_spanning_forest}(a). 
  In $E_1$, edges $(\mathsf{n_2}, \mathsf{n_5})$, 
  $(\mathsf{n_3}, \mathsf{n_6})$ and $(\mathsf{n_4}, \mathsf{n_5})$ are 
  non-tree edges while
  all other edges are tree edges.
\end{example}

We first look at update operations that involve non-tree edges, which
is the simpler case, and then move on to updates of tree edges. When
we delete a non-tree edge
$(u,v)$ in a connected component
$C(V',E')$, this does not affect the spanning tree and we do not have
to make any changes to it (we know that all vertices in
$C$ are still connected via the tree edges). Even better, if the
spanning tree is an (optimal) BFS-tree, it will 
remain an (optimal) BFS-tree, since 
taking away an edge from 
$C$ does not add any shortcuts between nodes that could lead to a better tree.

Inserting a new non-tree edge $(u,v)$, i.e., both, $u$ and $v$, are in
the same component $C$, means that the current spanning forest for $G$
is still valid. So, if we are only interested in maintaining spanning
trees for the components of $G$, we would not have to modify
anything. However, inserting a non-tree edge can invalidate 
that a spanning tree is a BFS-tree.
Assume that
$depth(u) + 1 < depth(v)$, then $v$ (and possibly some of its
ancestors) can be reached faster through $u$ than taking the existing
path from $v$ to the root of the tree.  We can fix this case.
We define $\Delta = depth(v) - depth(u)$. We
disconnect $v$ and $(\Delta - 2)$ of its ancestors ($v$'s
$(\Delta - 2)$-nd ancestor and $v$ have a distance of
$(\Delta - 2)$) from the spanning tree, reroot this subtree to make
$v$ the new root, and connect this subtree to $u$. The edge $(u,v)$
becomes a tree edge, while the edge previously connecting the
$(\Delta - 2)$-nd ancestor to the tree becomes a non-tree edge. We now
have a spanning tree 
that is a BFS-tree again. Note that the heuristic does not
guarantee the optimality of the BFS-tree.

\begin{example}
  Figure~\ref{fig:restore_BFS} shows an example of restoring a BFS-tree
  after inserting a non-tree edge ($\mathsf{n}_{5}$,
  $\mathsf{n}_{8}$). $\mathsf{n}_{5}$ can reach root $\mathsf{n}_{1}$
  faster through $\mathsf{n}_{8}$. Since $depth(\mathsf{n}_{8})+1$ $<$
  $depth(\mathsf{n}_{5})$, $\Delta$ = $depth(\mathsf{n}_{5})$ $-$
  $depth(\mathsf{n}_{8})$ = $4 - 1$ = 3, and $\Delta-2$ = 1, the
  $(\Delta - 2)$-nd ancestor of $\mathsf{n}_{5}$ is $\mathsf{n}_{4}$.
  We disconnect $\mathsf{n}_{4}$ from the tree, turning
  $\mathsf{n}_{5}$ into the root of the subtree and connecting this
  subtree to $\mathsf{n}_{8}$.  The previous tree edge
  ($\mathsf{n}_{3}$, $\mathsf{n}_{4}$) becomes a non-tree edge (not
  shown in Figure \ref{fig:restore_BFS}) and ($\mathsf{n}_{5}$,
  $\mathsf{n}_{8}$) becomes a tree edge. While the tree in
    Figure~\ref{fig:restore_BFS}(b) is a BFS-tree, it is not the BFS-tree with
    the optimal $S_d$ anymore. In Section~\ref{sec:maintaining_st} we 
    show how to improve $S_d$.

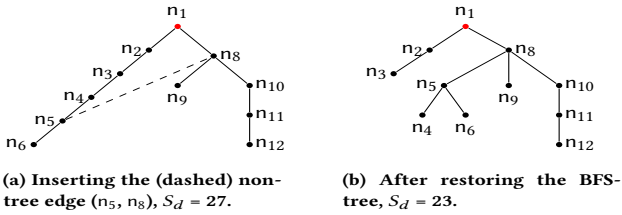
\begin{figure}[htb]\centering
\resizebox{3.8cm}{2.8cm}{
  \begin{subfigure}[b]{0.45\columnwidth}\centering
    \begin{tikzpicture}
      \tikzstyle{every node}=[fill=black,circle,inner sep=0pt,minimum size=2.5pt,]
      \node [fill=red, label=above:$\mathsf{n}_{1}$] (1) at (-4.0, 0.0) {};
      \node [label=left:$\mathsf{n}_{2}$] (2) at (-4.4, -0.4) {};
      \node [label=left:$\mathsf{n}_{3}$] (3) at (-4.8, -0.8) {};
      \node [label=left:$\mathsf{n}_{4}$] (4) at (-5.2, -1.2) {};
      \node [label=left:$\mathsf{n}_{5}$] (5) at (-5.6, -1.6) {};  
      \node [label=left:$\mathsf{n}_{6}$] (6) at (-6.0, -2.0) {};  
        
      \node [label=right:$\mathsf{n}_{8}$] (8) at (-3.5, -0.5) {};
      \node [label=below:$\mathsf{n}_{9}$] (9) at (-4.0, -1.0) {};
      \node [label=right:$\mathsf{n}_{10}$] (10) at (-3.0, -1.0) {};
      \node [label=right:$\mathsf{n}_{11}$] (11) at (-3.0, -1.5) {};
      \node [label=right:$\mathsf{n}_{12}$] (12) at (-3.0, -2.0) {};
       
      \draw[-] (1) to (2);
      \draw[-] (2) to (3);
      \draw[-] (3) to (4);
      \draw[-] (4) to (5);
      \draw[-] (5) to (6);
      
      \draw[dashed] (5) -- (8)  node [below, midway, fill=none]{};
      
      \draw[-] (1) to (8);
      \draw[-] (8) to (9);
      \draw[-] (8) to (10);
      \draw[-] (10) to (11);
      \draw[-] (11) to (12);
    \end{tikzpicture}
    \caption{Inserting the (dashed) non-tree edge
    ($\mathsf{n}_{5}$, $\mathsf{n}_{8}$), $S_d$ = 27.}
  \end{subfigure}
  }
  \quad \quad
  \resizebox{3.8cm}{2.8cm}{
  \begin{subfigure}[b]{0.45\columnwidth}\centering
    \begin{tikzpicture}
      \tikzstyle{every node}=[fill=black,circle,inner sep=0pt,minimum size=2.5pt,]
      \node [fill=red, label=above:$\mathsf{n}_{1}$] (1) at (-4.0, 0.0) {};
      \node [label=left:$\mathsf{n}_{2}$] (2) at (-4.5, -0.4) {};
      \node [label=left:$\mathsf{n}_{3}$] (3) at (-5.0, -0.8) {};
      
      \node [label=right:$\mathsf{n}_{8}$] (8) at (-3.4, -0.4) {};
      \node [label=left:$\mathsf{n}_{5}$] (5) at (-4.3, -1.0) {};  

      \node [label=below:$\mathsf{n}_{9}$] (9) at (-3.4, -1.0) {};
      \node [label=right:$\mathsf{n}_{10}$] (10) at (-2.7, -1.0) {};
      \node [label=right:$\mathsf{n}_{11}$] (11) at (-2.7, -1.5) {};
      \node [label=right:$\mathsf{n}_{12}$] (12) at (-2.7, -2.0) {};
      
      \node [label=below:$\mathsf{n}_{4}$] (4) at (-4.6, -1.5) {};
      \node [label=below:$\mathsf{n}_{6}$] (6) at (-4.0, -1.5) {};
      
      \draw[-] (1) to (2);
      \draw[-] (2) to (3);

      \draw[-] (4) to (5);
      \draw[-] (6) to (5);
      \draw[-] (5) to (8);
      
      \draw[-] (1) to (8);
      \draw[-] (8) to (9);
      \draw[-] (8) to (10);
      \draw[-] (10) to (11);
      \draw[-] (11) to (12);
    \end{tikzpicture}
    \caption{After restoring the BFS-tree, $S_d$ = 23.}
  \end{subfigure}
  }
  \caption{Restoring the BFS-tree.}
  \label{fig:restore_BFS}
\end{figure}

\end{example}

Let us now turn to updates involving tree edges.  If we insert a new
edge $(u,v)$ into $G$ and discover that $u$ and $v$ are located in
different components, $C_1$ and $C_2$, respectively, then we need to
merge $C_1$ and $C_2$ into a single component $C_3$. Consequently, the
spanning trees $T_1$ and $T_2$ currently representing $C_1$ and $C_2$
also need to be merged into a single spanning tree $T_3$. This
involves rerooting one of the trees and connecting it to the
other. Assume that we make $v$ the new root of $T_2$, which, w.l.o.g.,
is the smaller tree, and then connect it via $(u,v)$ to $T_1$, making
$(u,v)$ a tree edge in $T_3$. If we start with trees that 
are BFS-trees, 
the part covered by $T_1$ will still be one and the
edge $(u,v)$ is on the shortest path to connect to vertices in $T_2$,
which may not be a BFS-tree anymore after the rerooting.  Essentially,
this limits the damage we do to 
the smaller
tree. Instead of rerooting $T_2$, we could run BFS on $T_2$ starting
at node $v$ (to recreate a BFS-tree) and then connect $u$ to
$v$. This entails costs of $O(|V_2| + |E_{T_2}|)$, compared to
$O(depth(v))$ for rerooting the tree. The performance is the reason
we opt for the rerooting, even though it does not guarantee an optimal 
BFS-tree (more details on the implementation in
Section~\ref{sec:ourapproach} and the impact on the performance in
Section~\ref{sec:experiments}).

When deleting a tree edge, the spanning tree $T$ for $C$ is split into
two trees $T_1$ and $T_2$. However, we do not know yet whether this
will also split component $C$.  If we can find a \emph{replacement
  edge} $(x,y) \in E' \setminus E_T$ among the non-tree edges in $C$
that reconnects $T_1$ and $T_2$, then we know that the vertices in $C$
are still connected. In this case, $(x,y)$ becomes a tree edge in the
new, rearranged spanning tree for $C$ and is handled like the
insertion of a tree edge as described above (i.e., we reroot the smaller tree
and attach it to the other one).  However, we may have more
than one replacement edge. In this case,
we choose the edge connecting to the node closest to the root of
the larger tree. This is the fastest way from the root of the larger
tree to the smaller tree.  If we cannot find a replacement
edge, we know that $C$ has been split into two connected components
$C_1$ and $C_2$ by the deletion of $(u,v)$. The two parts of the
original spanning tree, $T_1$ and $T_2$, then represent $C_1$ and
$C_2$, respectively. If the original tree $T$ is a BFS-tree,
then $T_1$ and $T_2$ will also be a BFS-tree
(albeit not necessarily an optimal one).
Deleting a tree edge
is the most complex operation, we take a detailed
look in the following section.  While a single edge
always suffices to reconnect spanning trees after a deletion, the
problem is finding this edge efficiently without searching through
large parts of $T_1$ and $T_2$.

\subsection{Searching for a Replacement Edge}
\label{sec:replacementedge}

A naive approach of searching for a replacement edge after a
deletion is to run DFS or BFS on the resulting trees
$T_1(V_1,E_{T_1})$ and $T_2(V_2,E_{T_2})$.  This is costly for graphs
containing large connected components
($O(|V_1|+|V_2|+|E_{T_1}|+|E_{T_2}|$) if implemented naively.  There
are some optimizations we can apply, though. We only need to search the
smaller of the two trees $T_1$ and $T_2$: a replacement
edge can be found from either direction. So, we could run the search
on $T_1$ and $T_2$ in an interleaved fashion and immediately stop once
we have completely traversed one of the trees (or have found a
replacement edge).  Alternatively, keeping track of the size of subtrees in a
spanning tree, we could always run the search on the smaller tree.

In our approach, we create and maintain spanning trees in a way
to increase the likelihood of an uneven split. We define the
\emph{cut number} of an edge $e \in E_T$ in a tree $T(V',E_T)$, which
is the size of the smaller tree after splitting $T$ along $e$.

\begin{definition}[Cut number]
  \label{def:cutnumber}
  Given a tree $T(V',E_T)$ and an edge $e \in E_T$, we split $T$ into
  two subtrees, $T_1$ and $T_2$, by removing $e$ (every edge in a tree is a
  cut edge). We define the \emph{cut number} of $e$ as the size of the smaller
  tree: $c(e) = \min(|T_1|,|T_2|)$. Let $S_c(T) = \sum_{e \in E_T} c(e)$ be the
  sum of cut numbers for $T$. 
\end{definition}

The search for a replacement edge after deleting a tree edge is
proportional to the cut number of the edge we are deleting. Thus,
assuming a uniform distribution for selecting a cut edge, the average
costs of the search are equal to $\sfrac{S_c(T)}{|E_T|}$. These costs
are minimized for spanning trees that minimize $S_c$, as $|E_T|$ is
constant for any given spanning tree.

It is hard to analyze the cut number as defined in
Definition~\ref{def:cutnumber}, as we are summing over
minimums. However, there is an alternative way to compute the cut
number. We first formulate the following theorem (taken from
\cite{dobrynin2001wiener,Zelinka68}), which we use for computing the
cut number.

\begin{theorem}[Centroid and size of subtrees]
  \label{theo:centroid}
  Let $m$ be (one of) the centroid(s) of a tree $T(V',E_T)$.  Removing
  this centroid from the tree will create a forest consisting of trees
  $T_1, T_2, \dots, T_k$. For every tree $T_i$, $1 \leq i \leq k$,
  $|T_i| \leq \sfrac{|T|}{2}$, i.e., each tree $T_i$ contains at most
  half of the vertices of $T$.
\end{theorem}

Before computing the cut number of a tree, we move the root of the
tree to (one of) the centroid(s) $m$.
This allows us to get rid of the minimum in $S_c$, as we know
that every subtree connected to $m$ contains at most half of the
vertices. 
W.l.o.g. let $p_v$ be the parent of $v$,  
we go through all the edges $(p_v,v) \in E_T$.
Due to Theorem~\ref{theo:centroid}, we know that the cut number of
$(p_v,v)$ is equal to $size(v)$, the size of the subtree rooted at
$v$. Therefore,
\begin{eqnarray}
  S_c(T) = \sum_{v \in V' \setminus m} size(v) \label{eq:cutnumber}
\end{eqnarray}

\begin{lemma} \label{lemma:sc_sd} For a tree $T(V',E_T)$ whose root
  $r$ is a centroid, the sum of cut numbers, $S_c(T)$, is equal to the
  sum of distances, $S_d(T)$.
\end{lemma}

\inreport{
  \begin{proof}
    Shown in Appendix \ref{lemma_sc_sd_proof}
  \end{proof}
}

Thus, the sums $S_c$ and $S_d$ are directly related to each
other. Even better, utilizing Lemma~\ref{lemma:sc_sd} and
Equation~(\ref{eq:cutnumber}) (see Section~\ref{sec:ourapproach} for
details), we can maintain a low value for $S_c$ and $S_d$ using
information about the size of subtrees, which is much easier to
maintain in a dynamic spanning tree than information about the depth
of nodes.

With the next lemma we show that the BFS-spanning-tree $T_m$ with the
minimal sum of distances $S_d$ for a component will always have a
centroid as a root. For $T_m$, the average costs for evaluating
connectivity queries and searching for a replacement edge are
minimized.

\begin{lemma}
  \label{lemma:bfscentroid}
  Let $P = \{\mbox{BFS-tree with root } v|v \in V'\}$ be the set of
  BFS-trees for component $C = (V', E')$.  Let $T_m(V_m,E_m) \in P$
  with root $r$ being the BFS-tree in $P$ with minimal overall
  $S_d$ for all trees in $P$. Then $r$ is a centroid of $T_m$.
\end{lemma}

\inreport{
  \begin{proof}
  Shown in Appendix \ref{lemma_bfscentroid_proof}.
  \end{proof}
}

\subsection{Fixing Spanning Trees}
\label{sec:maintaining_st}

We have now identified what a spanning tree for a component has to
look like in the ideal case
to minimize the average costs for evaluating connectivity
queries and searching for a replacement edge: it is the BFS-tree with
the minimal sum of distances.  Next, we have a closer look at
how $S_d$ is affected by updates.  When we delete a non-tree edge in a
component, the value of $S_d$ for BFS-trees rooted at other nodes can
never decrease, as we now have fewer options to expand the search
frontier during BFS. So, we are on the safe side in this case.

\begin{figure}[htb]\centering
    \begin{tikzpicture}[scale=0.8]
      \tikzstyle{every node}=[fill=black,circle,inner sep=0pt,minimum size=2.5pt,]
      \node [fill=red, label=above:$\mathsf{n}_{8}$] (8) at (-4.0, 0.0) {};
      \node [label=left:$\mathsf{n}_{1}$] (1) at (-5.2, -0.5) {};
      \node [label=left:$\mathsf{n}_{5}$] (5) at (-4.4, -0.5) {};  
      \node [label=below:$\mathsf{n}_{9}$] (9) at (-3.6, -0.5) {}; 
      \node [label=right:$\mathsf{n}_{10}$] (10) at (-3.0, -0.5) {};
            
      \node [label=left:$\mathsf{n}_{2}$] (2) at (-5.6, -0.9) {};
      \node [label=left:$\mathsf{n}_{3}$] (3) at (-5.8, -1.3) {};

      \node [label=below:$\mathsf{n}_{4}$] (4) at (-4.2, -1.0) {};
      \node [label=below:$\mathsf{n}_{6}$] (6) at (-4.6, -1.0) {};
      
      \node [label=right:$\mathsf{n}_{11}$] (11) at (-2.6, -0.9) {}; 
      \node [label=right:$\mathsf{n}_{12}$] (12) at (-2.4, -1.3) {}; 
       
      \draw[-] (1) to (2);
      \draw[-] (2) to (3);

      \draw[-] (8) to (1);
      \draw[-] (8) to (5);
      \draw[-] (8) to (9);
      \draw[-] (8) to (10);
      
      \draw[-] (5) to (4);
      \draw[-] (5) to (6);
      \draw[-] (10) to (11);
      \draw[-] (11) to (12); 
    \end{tikzpicture}
  \caption{Restoring centroid property, $S_d$ = 18.}
  \label{fig:restore_centroid}
\end{figure}
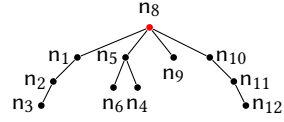

While inserting a non-tree edge and rearranging subtrees as described
in Section~\ref{sec:updating} keeps them BFS-trees, there might
now be a BFS-spanning-tree rooted at another vertex with a smaller
$S_d$. For example, assume that a connected component $C(V',E')$ only
contains the (solid) edges of tree $T(V',E_T)$ in
Figure~\ref{fig:restore_BFS}(a), i.e., $E' = E_T$. Then we insert the
(dashed) non-tree edge $(n_5,n_8)$ and restructure the tree to look as
depicted in Figure~\ref{fig:restore_BFS}(b). Clearly, this is a
BFS-tree. However, if we construct a spanning tree by running a BFS
starting from node $n_8$, we would get the tree $T'(V',E_{T'})$ shown
in Figure~\ref{fig:restore_centroid}, with
$S_d(T') = 18 < 25 = S_d(T)$. Running a BFS on (all) vertices of a
connected component after an insertion to find a BFS-tree with a
better value for $S_d$ is too expensive. Nevertheless, we can at least
restore the centroid property, i.e., if we notice that the root $r$ of the
current spanning tree is not a centroid, we reroot it. As we have seen
in Theorem~\ref{theo:centroid}, if we ever find a child $c_j$ of the
root with size greater than half of the vertices in the tree, we make
$r$ a child of $c_j$ and get a tree with a smaller sum of distances
$S_d$. While this does not guarantee the best overall spanning
tree for a component, it guarantees a tree that minimizes $S_d$
for all trees with root $c_j$ (see also
Definition~\ref{def:centroid}).

Ending up with a subtree that contains more than half of the vertices
can also happen during the insertion of a tree edge when we attach the
smaller to the larger tree.  Even splitting a spanning tree (in case
we do not find a replacement edge) can lead to this situation. For
example, if we delete edge ($n_2, n_3)$ in the tree shown in
Figure~\ref{fig:restore_BFS}(a) (before inserting $(n_5, n_8)$), we
end up with two BFS-spanning-trees, rooted at $n_3$ and $n_1$,
respectively, with a suboptimal $S_d$.  Since the spanning trees we
create tend to be flat with a high fan-out, going through all the
children of the root can take considerable time. Instead, we piggyback
the centroid restoration onto other operators.

Before we insert a tree or non-tree edge $(u,v)$, we have to go
to the root of the tree(s) containing $u$ and $v$, to find out whether
$(u,v)$ is a tree or non-tree edge. Thus, once we have reached the
root, we check whether the child we came through on our way to the
root has a size greater than one half of the size of the root after
the insertion. If this is the case, we make this child the new
root. Unfortunately, this does not work in the case of a deletion that
splits a connected component, as we do not necessarily pass through
the child at the root of the subtree containing more than half of the
nodes. Therefore, we also check the size of the child we navigate
through when we reach the root during the evaluation of a connectivity
query. This defers the restoration of the centroid. However, as long
as we do not have any connectivity query passing through this child,
this has no influence on the query costs.



\section{Implementing Spanning Trees}
\label{sec:ourapproach}

The implementation must be able to distinguish and handle tree
and non-tree edges (as defined in Definition~\ref{def:treeedges}) in
spanning trees.  We start out by defining the \emph{neighborhood} of a
vertex.

\begin{definition}[Neighborhoods]
  Given a connected component $C = (V',E')$, let $\Gamma_C(v)$ (with
  $v \in V'$) denote the \emph{neighborhood} of node $v$, i.e.,
  $\Gamma_C(v) = \{ u \in V' | (u,v) \in E' \}$ contains all nodes in
  $V'$ to which $v$ is directly connected.  Given a spanning tree
  $T = (V', E_T)$ for component $C$, the \emph{tree-edge neighborhood}
  $\Gamma_{C,T}^{te}(v) = \{ u \in V' | (u,v) \in E_T \}$ of node $v$
  is the set of nodes in $\Gamma_C(v)$ that are directly connected to
  $v$ via edges in $E_T$.  The \emph{non-tree-edge neighborhood}
  $\Gamma_{C,T}^{nte}(v) =\{ u \in V' | (u,v) \in E' \setminus E_T \}$
  of node $v$ contains all other edges in $\Gamma_C(v)$.  Thus,
  $\Gamma_C(v) = \Gamma_{C,T}^{te}(v) \cup \Gamma_{C,T}^{nte}(v)$.
\end{definition}

\begin{example}
  Consider component $C_1$ in Figure~\ref{fig:example_graph_G1}(a),
  the neighborhood of vertex $\mathsf{n_5}$, 
  $\Gamma_{C_1}(\mathsf{n_5}) = 
  \{\mathsf{n_2}, \mathsf{n_3}, \mathsf{n_4}\}$.  Given the
  corresponding spanning tree $T_1$ in Figure 
  \ref{fig:graph_spanning_forest}(a),
  the tree-edge neighborhood of node $\mathsf{n_5}$,
  $\Gamma_{C_1,T_1}^{te}(\mathsf{n_5})$ is $\{\mathsf{n_3}\}$, 
  while its non-tree-edge neighborhood 
  $\Gamma_{C_1,T_1}^{nte}(\mathsf{n_5})$ is 
  $\{\mathsf{n_2}, \mathsf{n_4} \}$.
\end{example}

\subsection{Dynamic Trees}
\label{sec:dynamictrees}

A \emph{dynamic tree} or \emph{D-tree} is a spanning tree with
additional information to facilitate its maintenance.

\begin{definition}[Dynamic tree (D-tree)] \label{def:dtree} A
  \emph{dynamic tree} (D-tree) for a spanning tree $T = (V', E_T)$ is
  a k-ary tree (with arbitrarily large $k$) in which each tree node
  has an attribute
  \begin{itemize}
  \item $key$, which acts as a unique identifier of a node
  \item $parent$, which is a pointer that links a node to its parent
  \item $children$, which is a set of pointers that connects a node to
    all its children
  \end{itemize}
  The attribute $key$ identifies each node. We store both, $parent$
  and $children$, as we need to navigate both ways, e.g. traversing
  via parents for connectivity queries and via children searching for
  a replacement edge.  We write $p(v)$ to denote a pointer to node
  $v$.

  We add two more attributes for efficiency reasons:
  \begin{itemize}
  \item attribute $size$ denoting the number of nodes found in the
    subtree rooted at a node.
  \item attribute $nte$ storing the non-tree edge neighborhood
    $\Gamma_{C,T}^{nte}$ of a node (as pointers to neighboring nodes).
  \end{itemize}
  Attribute $size$ plays a crucial role when minimizing $S_d$ and
  $S_c$ (cf. Section~\ref{sec:spanningtree}), while $nte$ allows us to
  embed the complete graph $G(V,E)$ into a D-tree forest. Not having
  to compute these attribute values on the fly speeds up the
  maintenance considerably. Adding an additional attribute to each 
  node to indicate which root it belongs to would speed up queries, 
  but at the price of slowing down updates. Every time we merge, split, 
  or reroot a spanning tree, we would have to update this attribute: 
  when merging or splitting we would need to update all the nodes in the 
  smaller tree and when rerooting all the nodes in the whole tree.
  
\end{definition}

\begin{figure}[htbp]\centering
  \subcaptionbox{D-tree $D_1$}{
  \resizebox{3.5cm}{3.0cm}{
    \begin{forest}
      dtree,
    [{$\mathsf{n}_{1}$}\\{s=6} \\ {nte=\{\}}, name=root,
      [{$\mathsf{n}_{2}$}\\{s=1} \\ {nte=\{$\mathsf{n}_{5}$\}}, dchild]
      [{$\mathsf{n}_{3}$}\\{s=2} \\ {nte=\{$\mathsf{n}_{6}$\}}, dchild
        [{$\mathsf{n}_{5}$}\\{s=1} \\ {nte=\{$\mathsf{n}_{2}, \mathsf{n}_{4}$\}}, dchild]
      ]
      [{$\mathsf{n}_{4}$}\\{s=2} \\ {nte=\{$\mathsf{n}_{5}$\}}, dchild
        [{$\mathsf{n}_{6}$}\\{s=1} \\ {nte=\{$\mathsf{n}_{3}$\}}, dchild]
      ]
    ]
    \fill[red] (root.parent anchor) circle[radius=2pt];
    \end{forest}
  }
  }
  \subcaptionbox{D-tree $D_2$}{
    \resizebox{2.5cm}{3.0cm}{
      
      \begin{forest}
        dtree,
      [{$\mathsf{n}_{9}$}\\{s=4} \\ {nte=\{\}}, name=root,
        [{$\mathsf{n}_{10}$}\\{s=2} \\ {nte=\{$\mathsf{n}_{11}$\}}, dchild,
          [{$\mathsf{n}_{12}$}\\{s=1} \\ {nte=\{\}}, dchild]
        ]
        [{$\mathsf{n}_{11}$}\\{s=1} \\ {nte=\{$\mathsf{n}_{10}$\}}, dchild
        ]
      ]
      \fill[red] (root.parent anchor) circle[radius=2pt];
      \end{forest}
  }}
  \caption{D-trees $D_1$ and $D_2$ for the spanning 
    trees $T_1$ and $T_2$ of Figure \ref{fig:graph_spanning_forest}, 
    respectively. We show $key$, 
    $size$ (abbreviated with $s$), and $nte$ as attributes, while
    $parent$ and $children$ are visualized using lines.}
  \label{fig:Dtrees}
\end{figure}
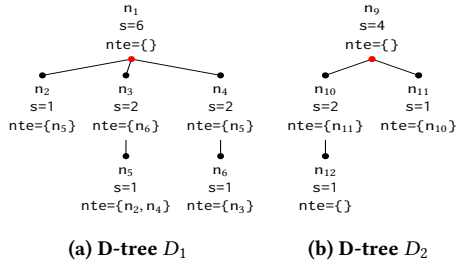

\begin{example}
  Figure \ref{fig:Dtrees} shows D-tree $D_1$ for the spanning tree
  $T_1$ in Figure \ref{fig:graph_spanning_forest}. Tree node
  $\mathsf{n_1}$ is the root (so $\mathsf{n_1}.parent$ $=$ $Null$),
  has three children ($\mathsf{n_1}.children$ $=$ $\{p(\mathsf{n_2})$,
  $p(\mathsf{n_3})$, $p(\mathsf{n_4})$\}) and no non-tree-edge
  neighbors ($\mathsf{n_1}.nte$ $=$
  $\Gamma_{C_1,T_1}^{nte}(\mathsf{n_1})$ $=$ $\{\}$.  The total number
  of nodes in the tree rooted at $\mathsf{n_1}$ is 6 (so,
  $\mathsf{n_1}.size$ $=$ 6). The edge $(\mathsf{n_2}, \mathsf{n_5})$
  is an example of a non-tree edge and is stored in the
  $nte$-attributes of nodes $\mathsf{n_2}$ and $\mathsf{n_5}$
  ($\mathsf{n_2}.nte$ $=$ $\{\mathsf{n_5}\}$ and $\mathsf{n_5}.nte$
  $=$ $\{\mathsf{n_2}\}$).
\end{example}

The attributes $parent$ and $children$ capture the tree-edge
neighborhood of a node:
$\Gamma_{C,T}^{te}(v) = \{ v.parent \cup v.children \}$ (we use the
dot notation to access attributes) while the non-tree-edge
neighborhood of a node is stored in attribute $nte$. Embedding the
complete graph $G(V,E)$ in a D-tree forest means that every vertex
$v \in V$ appears as a node $n_v$ in a D-tree (in the following, we
use $v$ and $n_v$ interchangeably) and every edge $(u, v) \in E$
appears in the set:
$\{(u,x) | x \in (u.parent \cup u.children \cup u.nte)\}$.


\subsection{Auxiliary Operations}
\label{sec:auxiliary_ops}

Before going into the details of the D-tree operations, we introduce
auxiliary operations to modify D-trees. These are needed, for example,
to prepare the merging of D-trees or to restore 
BFS-trees or the centroid property.  
The first auxiliary operation, shown in
Algorithm~\ref{alg:reroot}, is reroot. The reroot operation makes
$n_w$ the new root, which results in a new D-tree.  It follows the
path from the new root $n_w$ to the previous root, swaps the
parent/child relationship of two neighboring nodes, and updates the
$size$-attributes of the visited nodes.  

\begin{algorithm2e}[htb]
  \small
  \caption{reroot($n_w$)}
  \label{alg:reroot}
  \SetKwInOut{Input}{input}
  \SetKwInOut{Output}{output}
  
  \Input{tree node $n_w$ of D-tree with the root $r$}
  \Output{$n_w$, new root of the rerooted D-tree}

  $ch = n_w$; $cur = n_w.parent$; $n_w.parent = NULL$;\\
  \While{$cur \neq NULL$}{
    $g = cur.parent$\\
    $cur.parent = ch$ \\
    remove $ch$ from $cur.children$ \\
    add $cur$ to $ch.children$\\
    $ch = cur$; $cur = g$;\\
  }
  \While{$ch.parent \neq NULL$}{
    $ch.size$ = $ch.size$ - $ch.parent.size$ \\
    $ch.parent.size$ = $ch.parent.size$  + $ch.size$\\
    $ch = ch.parent$\\
  }
  \Return $u_w$
\end{algorithm2e}

\begin{example} \label{exam:reroot} In
  Figure~\ref{fig:reroot_example}, we employ reroot($\mathsf{n}_{1}$)
  on a D-tree and show the D-tree after the reroot operation.
\end{example}

\begin{figure}[htb]
  \scalebox{.50}{
    \begin{forest}
      dtree,
    [{$\mathsf{n_5}$}\\{s=9} , name=root,
      [{$\mathsf{n_9}$}\\{s=1}, dchild]
      [{$\mathsf{n_2}$}\\{s=7}, dchild,
        [{$\mathsf{n_6}$}\\{s=1}, dchild]
        [{$\mathsf{n_1}$}\\{s=5}, dchild,
          [{$\mathsf{n_3}$}\\{s=2}, dchild,
            [{$\mathsf{n_7}$}\\{s=1}, dchild]
          ]
          [{$\mathsf{n_4}$}\\{s=2}, dchild,
            [{$\mathsf{n_8}$}\\{s=1}, dchild]
          ]
        ]
      ]
    ]
    \fill[red] (root.parent anchor) circle[radius=2pt];
    \end{forest}
      \tikz[overlay, remember picture] \node (a) [xshift=-0.5cm,yshift=4cm] {};
    \tikz[overlay, remember picture] \node (b) [xshift=1.5cm,yshift=4cm] {};
    \tikz[overlay, remember picture] \draw[->,very thick] 
      (a) -- (b) node [above, midway]{$\mathsf{n}_{2}$ as the root};
  }
  \quad\quad
  \scalebox{.50}{
    \begin{forest}
      dtree,
      [{$\mathsf{n_2}$}\\{s=9}, name=root,
        [{$\mathsf{n_5}$}\\{s=2} , dchild,
          [{$\mathsf{n_9}$}\\{s=1}, dchild]
        ]
        [{$\mathsf{n_6}$}\\{s=1}, dchild]
        [{$\mathsf{n_1}$}\\{s=5}, dchild,
          [{$\mathsf{n_3}$}\\{s=2}, dchild,
            [{$\mathsf{n_7}$}\\{s=1}, dchild]
          ]
          [{$\mathsf{n_4}$}\\{s=2}, dchild,
            [{$\mathsf{n_8}$}\\{s=1}, dchild]
          ]
        ]
      ]
    ]
    \fill[red] (root.parent anchor) circle[radius=2pt];
    \end{forest}
      \tikz[overlay, remember picture] \node (a) [xshift=-0.5cm,yshift=4cm] {};
    \tikz[overlay, remember picture] \node (b) [xshift=1.5cm,yshift=4cm] {};
    \tikz[overlay, remember picture] \draw[->,very thick] 
      (a) -- (b) node [above, midway]{$\mathsf{n}_{1}$ as the root};
  }
  \quad\quad
  \scalebox{.50}{
    \begin{forest}
      dtree,
    [{$\mathsf{n}_{1}$}\\{s=9}, name=root,
      [{$\mathsf{n}_{2}$}\\{s=4}, dchild,
        [{$\mathsf{n_5}$}\\{s=2}, dchild,
          [{$\mathsf{n_9}$}\\{s=1}, dchild]
        ]
        [{$\mathsf{n_6}$}\\{s=1}, dchild]
      ]
      [{$\mathsf{n_3}$}\\{s=2}, dchild,
        [{$\mathsf{n_7}$}\\{s=1}, dchild]
      ]
      [{$\mathsf{n_4}$}\\{s=2}, dchild,
        [{$\mathsf{n_8}$}\\{s=1}, dchild]
      ]
    ]
    \fill[red] (root.parent anchor) circle[radius=2pt];
    \end{forest} 
  }
  \caption{Example of reroot operation. 
  The $nte$-attributes are not shown since they remain
the same.}
  \label{fig:reroot_example}
\end{figure}
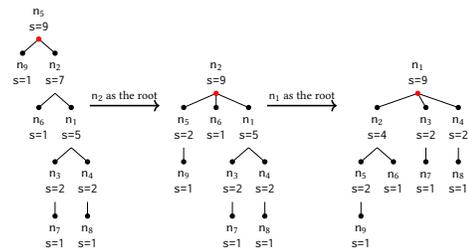

\begin{figure}[htb]
  \scalebox{.5}{
    \begin{forest}
      dtree,
    [{$\mathsf{n}_{1}$}\\{s=6}, name=root,
      [{$\mathsf{n}_{2}$}\\{s=1} , dchild]
      [{$\mathsf{n}_{3}$}\\{s=2}, dchild
        [{$\mathsf{n}_{5}$}\\{s=1}, dchild]
      ]
      [{$\mathsf{n}_{4}$}\\{s=2}, dchild
        [{$\mathsf{n}_{6}$}\\{s=1}, dchild]
        [{$\mathsf{n}_{10}$}\\{s=4}, name=root_sub, edge={thick, white}, 
          [{$\mathsf{n}_{12}$}\\{s=1}, dchild,
          ]
          [{$\mathsf{n}_{9}$}\\{s=2}, dchild,
            [{$\mathsf{n}_{11}$}\\{s=1}, dchild]
          ]
        ]
      ]
    ]
    \fill[red] (root.parent anchor) circle[radius=2pt];
    \fill[red] (root_sub.parent anchor) circle[radius=2pt];
  \end{forest}
  \quad
  \tikz[overlay, remember picture] \node (a) [xshift=-1.0cm,yshift=4cm] {};
  \tikz[overlay, remember picture] \node (b) [xshift=1.0cm,yshift=4cm] {};
  \tikz[overlay, remember picture] \draw[->,very thick] 
    (a) -- (b) node [above, midway]{update size};
  }
  \quad
  \scalebox{.5}{
    \begin{forest}
      dtree,
    [{$\mathsf{n}_{1}$}\\{s=10}, name=root,
      [{$\mathsf{n}_{2}$}\\{s=1}, dchild]
      [{$\mathsf{n}_{3}$}\\{s=2}, dchild
        [{$\mathsf{n}_{5}$}\\{s=1}, dchild]
      ]
      [{$\mathsf{n}_{4}$}\\{s=6}, dchild
        [{$\mathsf{n}_{6}$}\\{s=1}, dchild]
        [{$\mathsf{n}_{10}$}\\{s=4}, dchild, 
          [{$\mathsf{n}_{12}$}\\{s=1}, dchild,
          ]
          [{$\mathsf{n}_{9}$}\\{s=2}, dchild,
            [{$\mathsf{n}_{11}$}\\{s=1}, dchild]
          ]
        ]
      ]
    ]
    \fill[red] (root.parent anchor) circle[radius=2pt];
  \end{forest}
  \quad
    \tikz[overlay, remember picture] \node (a) [xshift=-1.0cm,yshift=4cm] {};
    \tikz[overlay, remember picture] \node (b) [xshift=1.0cm,yshift=4cm] {};
    \tikz[overlay, remember picture] \draw[->,very thick] 
      (a) -- (b) node [above, midway]{reroot($\mathsf{n}_{4}$)};
  }
  \quad
  \scalebox{.5}{
    \begin{forest}
      dtree,
    [{$\mathsf{n}_{4}$}\\{s=10}, name=root,
      [{$\mathsf{n}_{1}$}\\{s=4}, dchild,
        [{$\mathsf{n}_{2}$}\\{s=1}, dchild]
        [{$\mathsf{n}_{3}$}\\{s=2}, dchild
          [{$\mathsf{n}_{5}$}\\{s=1}, dchild]
        ]
      ]
      [{$\mathsf{n}_{6}$}\\{s=1}, dchild]
      [{$\mathsf{n}_{10}$}\\{s=4}, dchild, 
        [{$\mathsf{n}_{12}$}\\{s=1}, dchild,
        ]
        [{$\mathsf{n}_{9}$}\\{s=2}, dchild,
          [{$\mathsf{n}_{11}$}\\{s=1}, dchild]
        ]
      ]
      ]
    ]
    \fill[red] (root.parent anchor) circle[radius=2pt];
  \end{forest}
  }

  \caption{Example of link($\mathsf{n}_{4}$, $\mathsf{n}_{1}$, 
  $\mathsf{n}_{10}$). 
  The $nte$-attributes are not shown since they remain
  the same.}
  \label{fig:link_exam}
\end{figure}
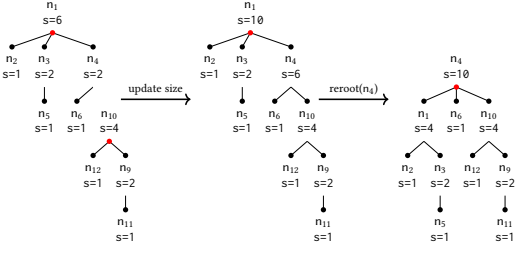

The link operation (\inreport{see Algorithm~\ref{alg:link}} {see
  technical report~\cite{technicalreport}} for pseudocode) takes two
D-trees that are currently not connected and connects them via a new
tree edge between $n_u$ (an arbitrary node in one of the D-trees) and
$n_v$ (the root of the other D-tree).\footnote{This means, that we may
  have to call a reroot operation on one of the trees before linking
  them.} During the linking, the $size$-attributes of the nodes on the
path from $n_u$ to $r_u$ are increased by $n_u.size$\inreport{(line
  \ref{alg:link_increment})}.  If we encounter a node on the path from
$n_u$ to the root that contains more than half of the nodes in the
merged tree\inreport{ (line \ref{alg:link_newroot})}, we restore the
centroid property (cf. Section~\ref{sec:maintaining_st}).

\makeatletter
\patchcmd{\@algocf@start}
  {-1.5em}
  {0pt}
  {}{}
\makeatother

\begin{example} \label{exam:link} Figure \ref{fig:link_exam} shows the
  operation link($\mathsf{n}_{4}$, $\mathsf{n}_{1}$,
  $\mathsf{n}_{10}$) that attaches $D_2$ (see Figure
  \ref{fig:Dtrees}(b)) to $D_1$ (see Figure \ref{fig:Dtrees}(a)).
  Values of $size$-attributes of nodes on the path from
  $\mathsf{n}_{4}$ to $\mathsf{n}_{1}$ are increased by
  $\mathsf{n}_{10}.size$ $=$ 4. Since $\mathsf{n}_{4}$ contains more
  than half of the nodes of the merged tree, $\mathsf{n}_{4}$ becomes
  the new centroid and we perform a reroot($\mathsf{n}_{4}$)
  operation.
\end{example}

The unlink operation (\inreport{see Algorithm~\ref{alg:unlink}} {see
  technical report~\cite{technicalreport} for pseudocode}) splits a
D-tree $D$ into two parts, by removing the tree edge between node
$n_v$, which is a non-root node in $D$, and its parent node. The
$size$-attributes of all (former) ancestors of $n_v$ are decreased by
$n_v.size$. After unlinking, $n_v$ becomes the root of a separate
D-tree, no adjustments are necessary in this tree.
For example, in Figure \ref{fig:deletete_example}(a), the
unlink($\mathsf{n_4}$) operation on $D_1$ of Figure \ref{fig:Dtrees}
results in two D-trees.

\subsection{Connectivity Queries}

Algorithm~\ref{alg:connquery} shows the pseudocode for running a
connectivity query $conn(n_u, n_v)$. As discussed in 
Section \ref{sec:maintaining_st}, this includes restoring the centroid
property (line \ref{alg:conn_reroot1} and line \ref{alg:conn_reroot2}).

\begin{algorithm2e}[htb]
  \DontPrintSemicolon
  \small
  \caption{conn($n_u$, $n_v$)}
  \label{alg:connquery}
  \SetKwInOut{Input}{input}
  \SetKwInOut{Output}{output}

  \Input{Tree nodes $n_u$ and $n_v$}
  \Output{True if $n_u$ and $n_v$ are connected, False otherwise}

  $d_u = Null$ \\
  \lWhile{$n_u.parent \neq Null$} {
    $d_u = n_u$; 
    $n_u = n_u.parent$
  }
  \lIf{$d_u \neq Null$ and  $d_u.size > n_u.size / 2$} {
    $n_u$ = reroot($d_u$) \label{alg:conn_reroot1}
  }

  $d_v = Null$ \\
  \lWhile{$n_v.parent \neq Null$} {
    $d_v = n_v$ ;
    $n_v = n_v.parent$
  }

  \lIf{$d_v \neq Null$ and $d_v.size > n_v.size / 2$} {
    $n_v$ = reroot($d_v$) \label{alg:conn_reroot2}
  }
  
  \Return $n_u.key == n_v.key$
\end{algorithm2e}

%
%

\subsection{Operations on Non-tree Edges}
\label{sec:operations_nte}

First, we determine if we are deleting a tree edge or a non-tree edge.
Consider an edge $(u,v) \in E'$ in a connected component
$C = (V',E')$.  If $u$ and $v$ are in a parent/child relationship in
the D-tree representing $C$, $(u,v)$ is a tree edge (which we cover in
Section~\ref{sec:deletete}), otherwise it is a non-tree edge (and,
thus, $u \in v.nte$ and $v \in u.nte$).

\subsubsection{Deleting Non-tree Edges}

Deleting a non-tree edge is the simplest update operation, as it does
not affect the structure of the spanning tree, we merely need to
update the $nte$-attributes of the corresponding nodes. \inreport{
  Algorithm~\ref{alg:deletente} shows the pseudocode for the deletion
  of a non-tree edge}{ The pseudocode for the deletion of a non-tree
  edge is available in the technical report~\cite{technicalreport}}.

\subsubsection{Inserting Non-tree Edges}

When inserting a new edge $(u,v)$ ($u,v \in V$) into a graph $G(V,E)$,
we first run a connectivity query $conn(u,v)$. If it returns 'True',
then $u$ and $v$ are in the same component $C$ and we are inserting a
non-tree edge. Algorithm~\ref{alg:insertnte} shows the pseudocode of
inserting a new non-tree edge (for details, see Section
\ref{sec:updating}).  The algorithm first determines the depths of
$n_u$ and $n_v$ and the root of $D$. If the difference of the depths
is less than two, we just add $(n_u,n_v)$ as a non-tree edge to
$D$. Otherwise, (w.l.o.g, assume that $depth(n_u) < depth(n_v)$), we
select the $(\Delta - 2)$nd ancestor of $n_v$ and unlink this ancestor
from $D$ (line \ref{alg:insertnte_unlink}); we make $h = n_v$ the root
of the resulting subtree and link this subtree to $D$ (line
\ref{alg:insertnte_link}).

\begin{algorithm2e}[htb]
  \small
  \caption{insert$_{nte}$($n_u$, $n_v$, $r$)}
  \label{alg:insertnte}
  \SetKwInOut{Input}{input}
  \SetKwInOut{Output}{output}

  \Input{Tree nodes $n_u$ and $n_v$ (in the same D-tree $D$), $r$ is root of $D$}
  \Output{Updated D-tree after insertion of non-tree edge $(n_u,n_v)$}

  determine $depth(n_u)$, $depth(n_v)$, and root $r$ of $D$ \\
  \lIf{$depth(n_u) \leq depth(n_v)$} {
    $l = n_u$; $h = n_v$
  } \lElse {
    $l = n_v$; $h = n_u$
  }
  $\Delta = depth(h) - depth(l)$ \\
  \If{$\Delta < 2$} {
    add $n_v$ to $n_u.nte$ \\
    add $n_u$ to $n_v.nte$ \\
    \Return $r$ 
  } \Else {
    $i = h$ \\
    \lFor{$x = 1$ \KwTo $\Delta - 2$} {
      $i = i.parent$
    }
    add $i$ to $i.parent.nte$ \\
    add $i.parent$ to $i.nte$ \\
    unlink($i$) \label{alg:insertnte_unlink}\\
    \Return link($l$, $r$, reroot($h$)) \label{alg:insertnte_link}
  }
\end{algorithm2e}

\subsection{Operations on Tree Edges}
\label{sec:operations_te}

\subsubsection{Inserting Tree Edges}
\label{sec:insertte}

We first discuss insertions of tree edges, which connect two
previously unconnected D-trees. This means, that the connectivity
query $conn(n_u,n_v)$ came back
with the result 'False'. We also know the roots of the trees
containing $n_u$ and $n_v$ now: they are $r_u$ and $r_v$,
respectively. Algorithm~\ref{alg:insertte} shows the pseudocode for
inserting the tree edge $(n_u,n_v)$ (details in Section
\ref{sec:updating}). Basically, we take the smaller
tree (w.l.o.g. assume that this is the tree containing $n_u$), reroot
it to $n_u$, and connect it to $n_v$. If necessary, the link operation also
restores the centroid property.

\begin{algorithm2e}[htb]
  \small
  \caption{insert$_{te}$($n_u$, $n_v$, $r_u$, $r_v$)}
  \label{alg:insertte}
  \SetKwInOut{Input}{input}
  \SetKwInOut{Output}{output}

  \Input{Tree nodes $n_u$ and $n_v$ and the roots $r_u$ and $r_v$ of the
    D-trees containing them}
  \Output{Merged D-tree after insertion of tree edge $(n_u,n_v)$}

  \lIf{$r_u.size < r_v.size$} {
    \Return link($n_v$, $r_v$, reroot($n_u$))
  } \lElse {
    \Return link($n_u$, $r_u$, reroot($n_v$))
  }
\end{algorithm2e}

\begin{example}
  Example for an insertion, insert$_{te}$($\mathsf{n_4}$,
  $\mathsf{n_{10}}$, $\mathsf{n_1}$, $\mathsf{n_{10}}$), can be seen
  in Example \ref{exam:link}. When inserting the tree edge
  ($\mathsf{n_4}$, $\mathsf{n_{10}}$), merging $D_1$ and $D_2$, we
  find that $D_2$ containing $\mathsf{n_{10}}$ has a smaller number of
  nodes. We conduct directly link($\mathsf{n_4}$, $\mathsf{n_1}$,
  $\mathsf{n_{10}}$) operation since $\mathsf{n_{10}}$ is already the
  root of the smaller tree, resulting the D-tree with $\mathsf{n_4}$
  as the centroid.
\end{example}

\subsubsection{Deleting Tree Edges}
\label{sec:deletete}

Algorithm~\ref{alg:deletete} shows the pseudocode for deleting tree
edges.  We first unlink the tree along the parent/child edge
$(n_u, n_v)$ and determine the root of the tree of the parent node
(the child node is the root of the unlinked subtree).  Next, we
conduct a BFS on the tree edges in the smaller tree (the one rooted at
$r_s$) to search for a replacement edge among the non-tree edges (line
\ref{alg:deletete_replacement}). If we do not find a replacement edge
(line \ref{alg:deletete_noreplacement}), we return the two unlinked
D-trees. We fix the centroid property of the smaller tree if it is
violated (line \ref{alg:deletete_centroid}).  If there are multiple
replacement edges, we pick one as described in
Section~\ref{sec:updating}.  In a replacement edge
$(n_{r_s}, n_{r_l})$, $n_{r_s}$ is located in the smaller tree created
by unlinking the input tree, while $n_{r_l}$ is located in the larger
tree (the one rooted at $r_l$).

\newcommand{\lIfElse}[3]{\lIf{#1}{#2 \textbf{else}~#3}}

\begin{algorithm2e}[htb]
  \DontPrintSemicolon
  \small
  \caption{delete$_{te}$($n_u$, $n_v$)}
  \label{alg:deletete}
  \SetKwInOut{Input}{input}
  \SetKwInOut{Output}{output}

  \Input{Nodes of $n_u$ and $n_v$ of deleted tree edge}
  \Output{Either reconnected D-tree if replacement edge is found or two
    separate D-trees otherwise}

  \lIfElse{$n_u = n_v.parent$} {
    $ch = n_v$
  }{
    $ch = n_u$ 
  }

  $(ch, r)$ = unlink($ch$) \\

  \lIfElse{$ch.size < r.size$} {
    $r_s = ch$; 
    $r_l = r$
  }{
    $r_s = r$;
    $r_l = ch$
  }
  $R = \{ (n_{r_s}, n_{r_l}) \, | \, n_{r_s} \! \in \! BFS(r_s) \, \wedge \,
  n_{r_l} \! \in \! n_{r_s}.nte 
  \, \wedge \, r_l \! \in \! anc(n_{r_l}) \}$ \label{alg:deletete_replacement} 

  \If{$R = \emptyset$} {\label{alg:deletete_noreplacement}
    \lIf{exists non-root $m$ with $m.size$ $>$ $\frac{r_s.size}{2}$}{ \label{alg:deletete_centroid}
      $r_s$ = reroot($m$)
    }
    \Return $(r_s, r_l)$
  } \Else {
    choose edge $(n_{r_s}, n_{r_l}) \in R$ with minimal $depth(n_{r_l})$  \\
    delete$_{nte}$($n_{r_s}$, $n_{r_l}$)\\
    \Return (insert$_{te}$($n_{r_s}$, $n_{r_l}$, $r_s$, $r_l$))
  }
\end{algorithm2e}

\begin{figure}[htbp]\centering
  \subcaptionbox{After unlink($\mathsf{n}_{4}$) \label{fig:unlink_exam}}{
  \scalebox{.48}{
    \begin{forest}
      dtree,
    [{$\mathsf{n}_{1}$}\\{s=4} \\ {nte=\{\}}, name=root,
      [{$\mathsf{n}_{2}$}\\{s=1} \\ {nte=\{$\mathsf{n}_{5}$\}}, dchild]
      [{$\mathsf{n}_{3}$}\\{s=3} \\ {nte=\{$\mathsf{n}_{6}$\}}, dchild
        [{$\mathsf{n}_{5}$}\\{s=1} \\ {nte=\{$\mathsf{n}_{2}$, $\mathsf{n}_{4}$\}}, dchild]
      ]
    ]
    \fill[red] (root.parent anchor) circle[radius=2pt];
    \end{forest}
    \begin{forest}
      dtree,
      [{$\mathsf{n}_{4}$}\\{s=2} \\ {nte=\{$\mathsf{n}_{5}$\}}, name=root,
        [{$\mathsf{n}_{6}$}\\{s=1} \\ {nte=\{$\mathsf{n}_{3}$\}}, dchild]
      ]
    \fill[red] (root.parent anchor) circle[radius=2pt];
    \end{forest}
  }}
  \subcaptionbox{After reroot($\mathsf{n}_{6}$)}{
  \scalebox{.48}{
    \begin{forest}
      dtree,
    [{$\mathsf{n}_{1}$}\\{s=4} \\ {nte=\{\}}, name=root,
      [{$\mathsf{n}_{2}$}\\{s=1} \\ {nte=\{$\mathsf{n}_{5}$\}}, dchild]
      [{$\mathsf{n}_{3}$}\\{s=3} \\ {nte=\{$\mathsf{n}_{6}$\}}, dchild
        [{$\mathsf{n}_{5}$}\\{s=1} \\ {nte=\{$\mathsf{n}_{2}$, $\mathsf{n}_{4}$\}}, dchild]
      ]
    ]
    \fill[red] (root.parent anchor) circle[radius=2pt];
    \end{forest}
    \begin{forest}
      dtree,
      [{$\mathsf{n}_{6}$}\\{s=2} \\ {nte=\{$\mathsf{n}_{3}$\}}, name=root,
        [{$\mathsf{n}_{4}$}\\{s=1} \\ {nte=\{$\mathsf{n}_{5}$\}}, dchild]
      ]
    \fill[red] (root.parent anchor) circle[radius=2pt];
    \end{forest}
  }}
  \subcaptionbox{After link($\mathsf{n_3}$, $\mathsf{n_1}$, $\mathsf{n_6}$)}{
  \scalebox{.50}{
    \begin{forest}
      dtree,
    [{$\mathsf{n}_{3}$}\\{s=6} \\ {nte=\{\}}, name=root,
      [{$\mathsf{n}_{1}$}\\{s=2} \\ {nte=\{\}}, dchild,
        [{$\mathsf{n}_{2}$}\\{s=1} \\ {nte=\{$\mathsf{n}_{5}$\}}, dchild]
      ]
      [{$\mathsf{n}_{5}$}\\{s=1} \\ {nte=\{$\mathsf{n}_{2}$, $\mathsf{n}_{4}$\}}, dchild]
      [{$\mathsf{n}_{6}$}\\{s=2} \\ {nte=\{\}}, dchild,
        [{$\mathsf{n}_{4}$}\\{s=1} \\ {nte=\{$\mathsf{n}_{5}$\}}, dchild]
      ]
    ]
    \fill[red] (root.parent anchor) circle[radius=2pt];
    \end{forest}
  }}
  \caption{Illustrations of 
  delete$_{te}$($\mathsf{n_1}$, $\mathsf{n_4}$) on D-tree $D_1$.}
  \label{fig:deletete_example}
\end{figure}
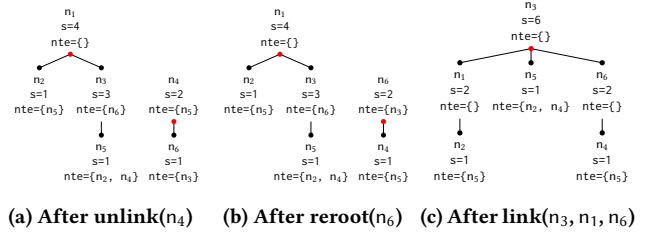

\begin{example}
  Figure \ref{fig:deletete_example} illustrates
  delete$_{te}$($\mathsf{n_1}$, $\mathsf{n_4}$) on $D_1$.  First, we
  remove the subtree rooted at $\mathsf{n_4}$ via
  unlink($\mathsf{n_4}$), creating two D-trees.  The D-tree with
  $\mathsf{n_4}$ as root is smaller in size, i.e., $r_s$ $=$
  $\mathsf{n_4}$ and $r_l$ $=$ $\mathsf{n_1}$.  We conduct a BFS
  starting at $\mathsf{n_4}$ to find replacement edges for the deleted
  tree edge ($\mathsf{n_1}$, $\mathsf{n_4}$) and get back
  $R= \{(\mathsf{n_4}, \mathsf{n_5}), (\mathsf{n_6}, \mathsf{n_3})\}$
  (line \ref{alg:deletete_replacement}).  We select the non-tree edge
  $(\mathsf{n_6}, \mathsf{n_3})$ as the replacement edge since the
  depth of $\mathsf{n_3}$ ($=$ 1) is smaller than the depth of
  $\mathsf{n_5}$ ($=$ 2).  We delete the non-tree edge
  $(\mathsf{n_6}, \mathsf{n_3})$, and run
  insert$_{te}$($\mathsf{n_6}$, $\mathsf{n_3}$, $\mathsf{n_4}$,
  $\mathsf{n_1}$).
\end{example}

Finally, we analyze the average case time complexity of the
operators. Deleting a non-tree edge $(u,v)$ is the simplest operation: we
just need to remove $u$ and $v$ from $v.nte$ and $u.nte$, respectively,
which takes constant time. The average cost for all auxiliary operations,
connectivity queries, and insertions of tree and non-tree edges is
proportional to the average distance between roots and all the other nodes, that is 
$\frac{S_d}{|V|}$, since all these operations involve traversing a spanning
tree from a node to a root. Deleting a tree edge requires the traversal of
the smaller tree and, potentially, the selection of a replacement edge. On
average, the cost for traversing the smaller tree is equal to the average
cut number, i.e., $\frac{S_c}{|V|}$. When determining whether a non-tree
edge is a replacement edge or not, we check if the node on the other side of
the edge belongs to the other tree, which has costs similar to a query.

\section{Experimental Evaluation}
\label{sec:experiments}

\subsection{Setup}
\label{sec:setup}

\textbf{Hardware and environment.}  All algorithms were implemented in
Python 3. The experiments were conducted on a single machine with
500GB RAM, running Debian 10.  All experiments were run
  10 times on the same machine, showing very similar results.

\textbf{Inserting and deleting edges.}  We start with empty graphs and
insert (and delete) edges one at a time. When inserting a new edge $e$
into the graph at time $t^e$, we assign a \emph{survival time} $t^e_d$
to $e$, i.e., the edge is deleted at time $t^e + t^e_d$. If $e$ is
re-inserted while still in the graph, e.g., at time $t^e_r$ (with
$t^e < t^e_r < t^e + t^e_d$), the survival of $e$ is extended, i.e.,
the deletion is rescheduled to $t^e_r + t^e_d$.  The deletion of edges
models that connections in graphs such as social or collaborative
networks become inactive after some time. Due to the different
granularity of time frames in the different graphs, we set $t^e_d$ to
five years for the Semantic Scholar (SC) dataset and to
fourteen days for all other datasets. 

\textbf{Setup of
  measurements.}  
Let $t_s$ and $t_e$ be the starting time and ending time
for all updates we run on the graph, respectively.  We examine
$test\_num$ snapshots, or testing points, of the spanning trees, which
are uniformly distributed in the period from $t_s$ to $t_e$.  We use
\emph{test\_frequency} $=$ $\sfrac{(t_e - t_s)}{test\_num}$ to define
how frequently we evaluate connectivity queries.  For all graphs
except SC, we set $test\_num = 100$, which means that every
$\sfrac{(t_e - t_s)}{100}$ steps, we run and evaluate connectivity
queries.  In the SC dataset, the edges are inserted on a yearly
basis, so we introduce a testing point every year.  For the timespan
$t_s$ to $t_e$, we accumulate the run time of all update operations
and show the average run time. There are variations 
in the size of the snapshots depending on the datasets. For example,
the size of the snapshots of the Tech and YT
datasets are close to the size of the actual dataset, while the snapshots
for the SC dataset reach the same order of magnitude as the actual dataset
toward the end of an experimental run.

\textbf{Evaluating connectivity queries.} At each testing point, we
run connectivity queries for all pairs of vertices in small graphs and
for 50 million uniformly distributed pairs in large graphs (as the
total number of pairs in large graphs becomes impractical). We
consider graphs with fewer than 10K vertices small graphs.

\subsection{Datasets}
\label{sec:datasets}

Every graph in our datasets is represented by a set of edges with
timestamps (one for the insertion time and another one for the
deletion time). All edges are undirected and we use $|V|$ and $|E|$ to
denote the number of vertices and edges for a graph, respectively.  We
use the following ten real-world graphs for our experimental studies.
\begin{table}[htb!]\centering
    \caption{Characteristics of datasets.}
    \label{table:datasets}
    \begin{tabular}{|c|c|c|c|}
    \hline
    Name & $|V|$ & $|E|$ & \# updates\\
    \hline
    \textbf{\small email-dnc (DNC)}{\scriptsize~\cite{nr}} & 1.9 $ \times 10^3$ & 3.74 $  \times 10^4$ & 3.2 $\times 10^4$\\
    \hline
    \textbf{\small Call (CA)}{\scriptsize~\cite{nr}}& 7 $ \times 10^3$ & 5.1 $  \times 10^4$ &  2.3 $\times 10^4$\\
    \hline
    \textbf{\small messages (MS)}{\scriptsize~\cite{nr}}& 2 $ \times 10^3$ & 6 $  \times 10^4$ & 6.3 $\times 10^4$\\
    \hline
    \textbf{\small FB-FORUM (FB)}{\scriptsize~\cite{nr}}& 8.99 $ \times 10^2$ & 3.4 $  \times 10^4$ & 3.8 $\times 10^4$\\
    \hline
    \textbf{\small Wiki-elec (WI)}{\scriptsize~\cite{nr}}& 7.1 $ \times 10^3$ & 1.07 $  \times 10^5$ & 2.1 $\times 10^5$\\
    \hline
    \textbf{\small tech-as-topology (Tech)}{\scriptsize~\cite{nr}}& 3.4 $ \times 10^4$ & 1.71 $  \times 10^5$ & 2.7 $\times 10^5$\\
    \hline
    \textbf{\small Enron (EN)}{\scriptsize~\cite{nr}} & 8.7 $ \times 10^4$ & 1.1$  \times 10^6$ & 1.28 $\times 10^6$ \\
    \hline
    \textbf{\small youtube-growth (YT)}{\scriptsize~\cite{mislove09}} & 3.2 $ \times 10^6$ & 1.44$  \times 10^7$ & 2.47 $\times 10^7$\\
    \hline
    \textbf{\small Stackoverflow (ST)}{\scriptsize~\cite{dataset_Stackoverflow}} & 2.6 $ \times 10^6$ & 6.3$  \times 10^7$ & 7$  \times 10^7$\\
    \hline
    \textbf{\small Semantic Scholar (SC)}
    {\scriptsize~\cite{ammar}} & 6.5 $ \times 10^7$ & 8.27$  \times 10^9$ & 9.36$  \times 10^9$\\
    \hline
    \end{tabular}
\end{table}

\subsection{Evaluated Methods}
\label{evaluated_methods}

We evaluate the performance of connectivity queries and maintenance
operations for the following methods:
\begin{itemize}
  \item our D-tree.
  
  \item $_n$D-tree, a naive version of Dtree, that neither maintains
    the BFS-tree nor the centroid property, which makes it easier (and
    faster) to update. A performance gap between $_n$D-trees and
    D-trees shows the effectiveness of the heuristics utilized in the
    D-tree.

  \item $opt$, optimal BFS tree: after each update, we run BFS over
    all vertices in the connected components affected by the update to
    determine the BFS-tree with minimal $S_d$.  This shows how much
    our D-tree deviates from the optimal case.

  \item ET-tree: maintains an Euler tour
    (ET)~\cite{tarjan1985efficient} of a spanning tree. To guarantee
    the worst-case behavior for connectivity queries, the ET is mapped
    to a balanced binary tree~\cite{David1997experiment, henzinger99},
    which means that an ET-tree is not a spanning tree anymore. As a
    consequence, update operations become more expensive (for details,
    see~\cite{henzinger99}). Many of the algorithms mentioned in
    Section~\ref{sec:related} are based on ET-trees, adding various
    optimizations to them~\cite{henzinger99, Holm2001,Thorup2000,
      wulff2013faster}.

  \item $HK$, the algorithm by Henzinger and
    King~\cite{henzinger1995randomized,henzinger99}, is also based on
    ET-trees, adding information -- in the form of a weight attribute
    -- about the number of non-tree edges in a subtree.  This allows
    the algorithm to terminate the search for a replacement edge early
    (if weight = 0 for a subtree). The early termination and a
    sampling scheme employed in the search achieves the reported
    amortized complexity. We implement $HK$ with one edge level, as
    Alberts et al. have shown that this version
    consistently outperforms the version with multiple
    levels~\cite{David1997experiment}.  $HK$ is the state-of-the-art
    algorithm, since this is the best algorithm among those with a
    worst-case guarantee mentioned in Section~\ref{sec:related} that
    has been fully implemented and evaluated empirically.  
    \item online BFS and DFS.
    \item Insertion-only algorithms: union-find algorithm
      ~\cite{Tarjan75, tarjan1983data} and DBL~\cite{DBL}.  

\end{itemize}

\subsection{Diameters of Real-world Graphs}

\begin{figure}[htb] \centering
  \begin{subfigure}[b] {0.4\columnwidth} 
    \resizebox{3.5cm}{1.9cm}{
      \includegraphics{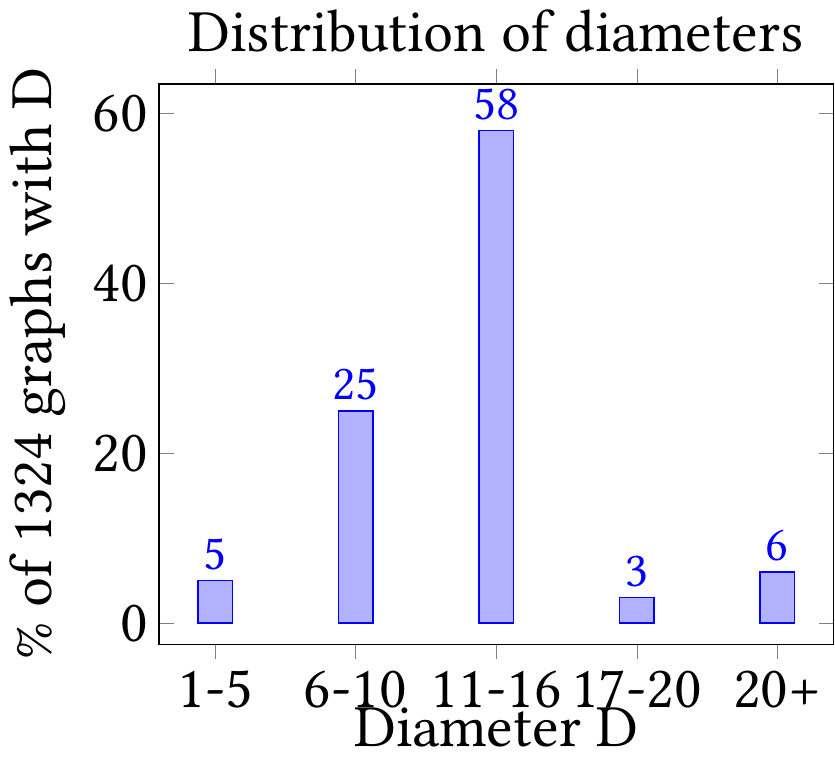}
    }  
    \caption{Distribution of diameters (89.5\% $\leq 16$).}
    \label{fig:diameter_dist}
  \end{subfigure}
  \quad
  \begin{subfigure}[b]{0.4\columnwidth} 
    \resizebox{3.5cm}{1.9cm}{
      \includegraphics{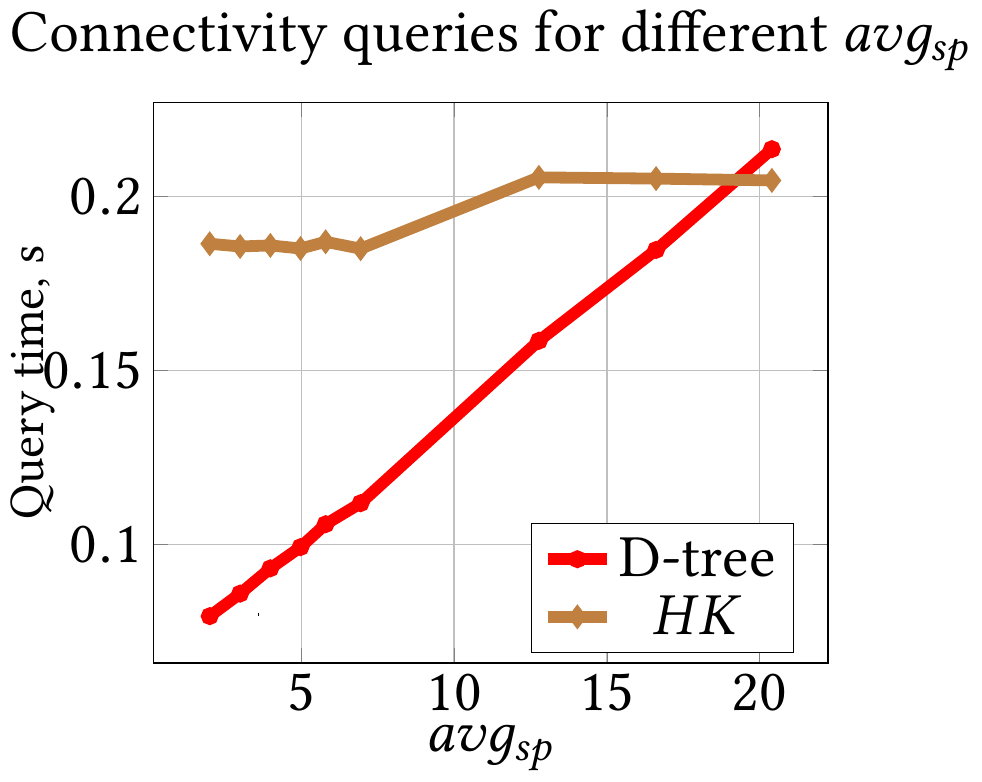}
    }
    \caption{D-tree outperforms $HK$ when $avg_{sp} \leq 16.6$.}  
    \label{fig:avg_sp}  
  \end{subfigure}
  \caption{Diameters for real-world graphs and $avg_{sp}$.}  
\end{figure}

Before comparing the different algorithmic approaches, we take a look at
  an important property of graphs and its impact on the performance of our
  D-tree, namely the diameter of graphs.
Algorithms guaranteeing worst-case performance for connectivity
queries, such as $HK$, focus on graphs with large diameters where the
benefits of their approach are most pronounced.  Dealing with
worst-case scenarios adds considerable overhead to those
algorithms. However, among 1324 real-world graphs we
investigated~\cite{KONECT_diameter} (see
Figure~\ref{fig:diameter_dist}), 1185, or 89.5\%, had a diameter not
larger than sixteen. For graphs with small diameters, we can easily
build and maintain D-trees with a high fanout and low depth (which is
bounded by the diameter of the graph), thus achieving very good
average-case performance for those graphs. This gives us an edge over
$HK$ in most real-world scenarios, as D-trees have a much higher
fanout than the balanced binary trees employed by $HK$.

We quantify the difference between D-trees and $HK$ by comparing their
connectivity query performance for different values of $avg_{sp}$, the
average sum of lengths of the shortest paths over all pairs of
vertices in a graph ($avg_{sp}$ is upper-bounded by the diameter). Let
$C = (V', E')$ be a connected component and $dist_C(u, v)$ the length
of the shortest path between $u \in V'$ and $v \in V'$,
\[
  avg_{sp}(C) = (\sum_{u<v}dist_C(u, v)) / \binom{|V'|}{2}.
\]

As $avg_{sp}$ (and the diameter) is expensive to compute for a given
graph, we generated synthetic graphs with a central node and $N=480$
other nodes arranged around this node. We connect $k$ line graphs,
each containing $\sfrac{N}{k}$ vertices, to the central node: this
regular structure allows us to compute $avg_{sp}$ (and the diameter)
more efficiently. Figure~\ref{fig:avg_sp} shows the connectivity query
performance of D-trees and $HK$ for different values of
$avg_{sp}$. D-trees outperform $HK$ for graphs with
$avg_{sp} \leq 16.6$, so we expect D-trees to outperform $HK$ for
\emph{at least} 89.5\% of the real-world graphs from
Figure~\ref{fig:diameter_dist}, due to the diameter being an upper
bound for $avg_{sp}$.

\subsection{Comparison with BFS/DFS}
\label{sec:BFSDFS}

We compared the runtime of connectivity queries for D-trees with that of
BFS/DFS, which acts as a baseline. The worst-case runtime complexity of
BFS/DFS is $O(|V| + |E|)$\cite{cormen2009introduction} and our experiments confirm that the runtime of
this approach is too high for practical purposes: on average, BFS/DFS is
several orders of magnitude slower than D-trees. For example, for one of the
smaller graphs, WI, running connectivity queries for all pairs of vertices,
which amounts to around 25 million queries, takes BFS/DFS more than eight
days to complete. In contrast, D-trees run this set of queries in 23
seconds. We ran the queries on the complete graph, i.e., we inserted all the
edges without deleting any. Clearly, BFS/DFS does not have any maintenance
costs, but it only took us 200ms to build the D-trees for the WI-graph from
scratch.

\subsection{Insertion-only Algorithms}
\label{sec:insertion_only}

\begin{figure}[htb!]\centering
  \includegraphics{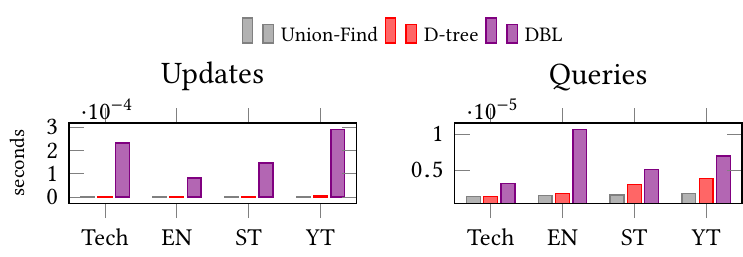}
\caption{Average run time for insertions and queries.}
\label{fig_insertiononly}
\end{figure}

Next, we compare D-trees with DBL and union-find~\cite{Tarjan75,
  tarjan1983data}, which is still considered the state-of-the-art
algorithm for insertion-only graphs~\cite{wulff2013faster}.
We measured the average query and insertion performance per operator
for D-trees, DBL, and union-find on the large graphs (excluding SC,
as DBL took too long to construct the 2-hop labeling). 
The left-hand side of
Figure~\ref{fig_insertiononly} shows the time for inserting all the
edges. Clearly, DBL is the slowest algorithm (even though we ran the
insertions in a batch, which adds the smallest overhead) and D-trees
are slightly slower than union-find. The right-hand side of
Figure~\ref{fig_insertiononly} shows the average runtime of running
50 million random connectivity queries (after inserting all the
edges in a first step). Unsurprisingly, union-find is the fastest
algorithm, followed by D-trees, and DBL comes in last again. DBL is
slow, because it needs to run BFS for the insertions and from time
to time also for queries. Although, union-find is the fastest
algorithm, it is not applicable to fully dynamic graphs. It does not
support deletions, as it only maintains compressed paths from nodes
to roots and does not preserve connections among non-root vertices.

\begin{figure*}[htb] \centering
  \begin{subfigure}{\textwidth}\centering
    \includegraphics{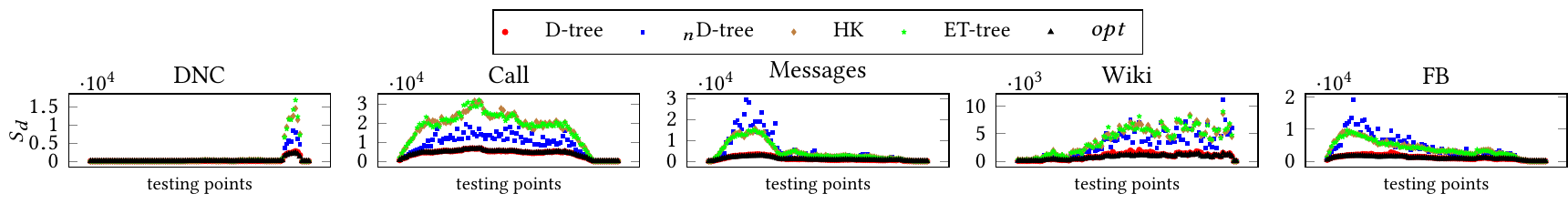}
  \end{subfigure}
  \begin{subfigure}{\textwidth}\centering
    \includegraphics{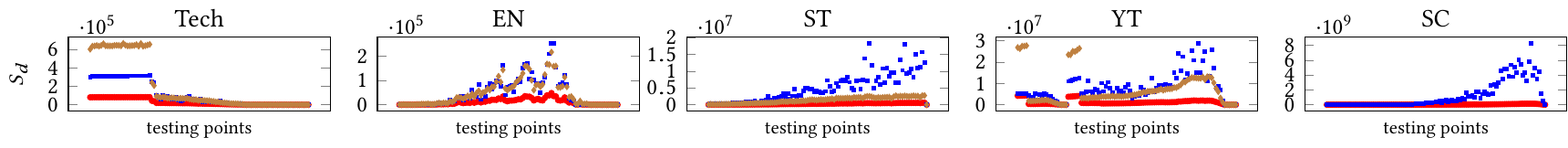}
  \end{subfigure}
  \caption{$S_d$ for spanning trees (forest) for graphs. }
  \label{fig:Sd}
\end{figure*}

\begin{figure*}[htb] \centering
  \begin{subfigure}{\textwidth}\centering
    \includegraphics{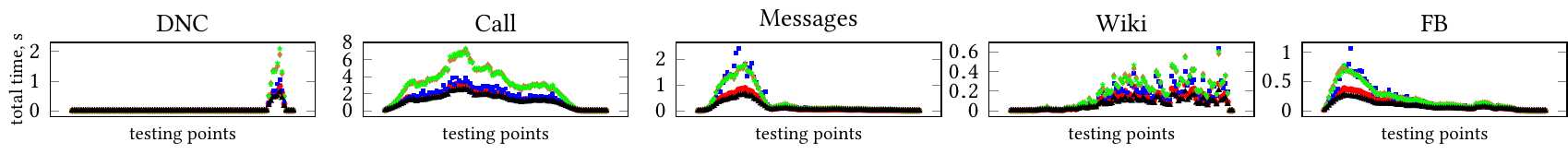}
  \end{subfigure}

  \begin{subfigure}{\textwidth}\centering
    \includegraphics{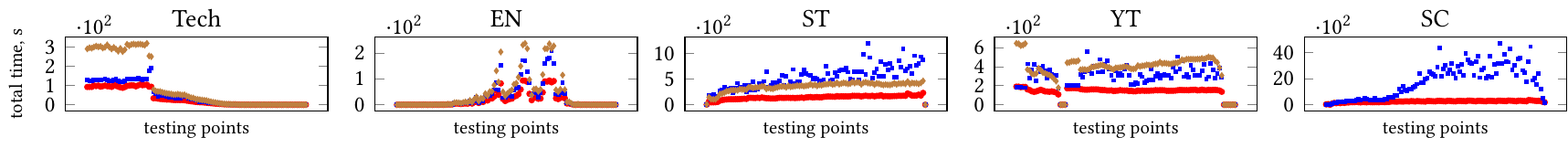}
  \end{subfigure}
  \caption{Query performance}
  \label{fig:query}
\end{figure*}

\begin{figure*} \centering
  \includegraphics{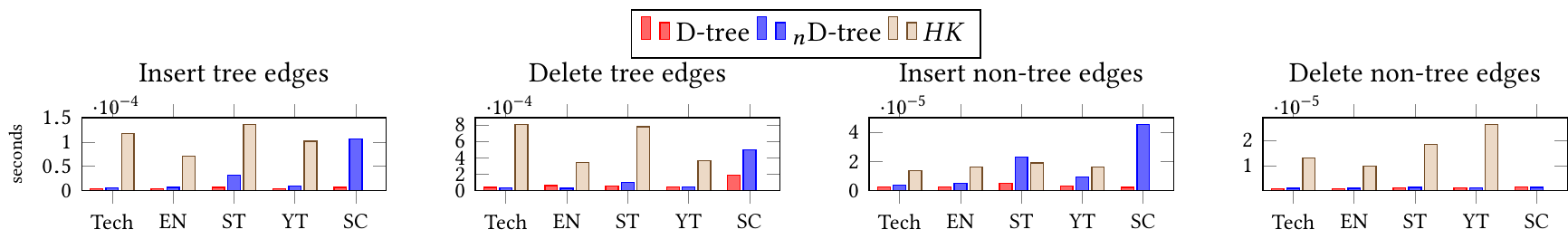}
  \caption{Average run time for updates.}
  \label{fig:operations_large}
\end{figure*}

\subsection{Distances between Roots and Nodes}

Here we confirm that the techniques we use for maintaining spanning
trees, namely preserving BFS-trees (if possible to do so
  efficiently), considering short-cuts when inserting non-tree edges,
  and re-establishing the centroid property, lead to small values for
  $S_d$.  In Figure~\ref{fig:Sd}, we show the value of $S_d$ for the
current spanning forest for every snapshot. The upper row depicts the
results for small graphs, for which we include the expensive methods
$opt$ and ET-tree. The best possible spanning forest is created by
$opt$, which computes the optimal BFS-tree.  We observe that our
D-tree is very close to $opt$ and much better than $_n$Dtree,
demonstrating the effectiveness of the heuristics for maintaining the
spanning forest. Our D-tree also has better values for $S_d$ than the
ET-tree and $HK$. The difference between the ET-tree and $HK$ is
minimal since both employ a treap~\cite{seidel1996randomized} to
balance the tree. The lower row of Figure~\ref{fig:Sd} shows the
results for large graphs and, again, our D-tree creates trees with
small $S_d$ values and is able to maintain the lead over time. We do
not show results for $opt$ and ET-trees for large graphs, as these
methods are very inefficient: $opt$ spends about 10 seconds per update
on the ST-graph (in contrast to less than one millisecond for D-trees)
and we do around 20 million updates in total per experiment; after a
couple of updates on the ST-graph, deletions on ET-trees are three
orders of magnitude slower than those on D-trees. We do not show
  results for $HK$ on the SC graph because $HK$ ran for fourteen days
  and was not able to finish in that time.


\inreport{Figure~\ref{fig:distributions} in the appendix gives a detailed 
insight into the distribution of
node depths in the various trees. On average, the nodes in our D-trees
are much closer to the roots. For small graphs (upper row of
Figure~\ref{fig:distributions}), we are very close to $opt$. For large graphs
(lower row of Figure~\ref{fig:distributions}), D-trees also outperform the other
methods.}{
Figure 16 in the technical report~\cite{technicalreport} 
gives a detailed insight into the distribution of
node depths in the various trees. On average, the nodes in our D-trees
are much closer to the roots. For small graphs (upper row of
Figure 16), we are very close to $opt$. For large graphs
(lower row of Figure 16), D-trees also outperform the other
methods.
}

\subsection{Performance for Connectivity Queries}

As we have shown in Theorem~\ref{theorem:average_cost}, the average
query costs are directly related to $S_d$. This is confirmed by our
experiments on query performance in Figure~\ref{fig:query}.  The
results are strongly correlated to those for $S_d$ in
Figure~\ref{fig:Sd}.  The average Pearson correlation between
  $S_d$ and query time over all datasets is 0.904842. The upper row
of Figure~\ref{fig:query} for small graphs demonstrates that the
performance of D-trees is very close to that of $opt$. Additionally,
D-trees consistently outperform $_n$D-trees, ET-trees, and $HK$ for
all graphs.  $avg_d$, the average distances between nodes and
  roots, is less than ten in D-trees while $avg_d$ for $HK$ is several
  times larger. 

\subsection{Performance for Update Operations}

Figure~\ref{fig:operations_large} shows the run times for update
operations. First, we see that $HK$ is much slower than the other
techniques (the differences are usually an order of magnitude). While
balanced binary trees offer good worst-case performance, they are much
deeper than D-trees. Moreover, $HK$ does not use spanning trees but a
more complex representation, adding to the overhead of update
operations. Next, we compare D-trees to $_n$D-trees to show the
effectiveness and costs of our heuristics. When deleting non-tree
edges, the differences are minimal: the overhead for preserving
BFS-trees 
in D-trees is very small. We observe the biggest differences for
inserting (tree and non-tree) edges. Since $_n$D-trees do not utilize
any heuristics for minimizing $S_d$, the distances between the roots
and other nodes in the spanning trees tend to grow over time. This has
a negative impact on insertions (and not just queries), because we
have to navigate to the roots of the spanning trees to determine
whether we insert a tree or non-tree edge. When deleting tree edges,
there is no clear winner between D-trees and $_n$D-trees. While
D-trees have a smaller cut number, they search through all potential
replacement edges to pick the best one (lowering $S_d$).  $_n$D-trees
terminate the search for a replacement edge as soon as they find the
first one.

\subsection{Discussion}
\label{sec:ex_discussion}

D-trees outperform HK in querying and inserting tree and non-tree
edges, because of the smaller $S_d$ in the D-trees. The ET-trees
employed by HK are shaped differently and do not represent spanning
trees directly. Basically, the occurrences of nodes in an Euler tour
of a spanning tree are mapped into a balanced binary tree such that
the in-order traversal of this tree is the Euler tour. This makes it
independent of the diameter of a graph and results in trees of depth
$\log_2(n)$ ($n$ being the number of nodes). Consequently, in the
worst case, a lookup on this tree is still logarithmic in the number
of nodes. However, it cannot take advantage of graphs with small
diameters, the nodes are embedded much deeper in the tree compared to
a D-tree. It gets even worse when deleting a non-tree edge: HK has
logarithmic runtime for this case (in contrast to the constant runtime
in D-trees).
On average, D-trees have very small cut numbers $S_c$, usually less
than fifteen, often smaller than ten. Due to the structure of the
ET-tree, the splits are more even, resulting in longer searches on
larger trees (usually more than an order of magnitude larger compared
to D-trees). Even though D-trees go through all non-tree edges when
searching for a replacement edge (while HK takes the first valid edge
it finds), due to the small $S_c$ and $S_d$, this is still efficient.

\section{Conclusion}

We identify two crucial parameters for optimizing connectivity queries
via spanning trees in fully dynamic graphs: $S_d$, the sum of
distances between nodes in a tree and its root, and $S_c$, the cut
number of a tree. Due to the high cost of maintaining trees that
  minimize $S_d$ and $S_c$, we develop a data structure, called
D-tree with heuristics to keep the values of $S_d$ and $S_c$ small
when updating the trees. This makes the evaluation of connectivity
queries and the maintenance of spanning trees more
efficient. Moreover, we show that it is possible to implement our
heuristics with a low overhead, i.e., we only need to know the size of
each subtree in a spanning tree. Extensive experiments with real-world
datasets demonstrate that our approach has a performance close
  to optimal BFS-trees and outperforms algorithms that guarantee
worst-case complexity. For instance, maintaining D-trees is up to
fifty times faster than $HK$ and D-trees have a much better average
query performance.

For future work, we plan to extend our approach for connectivity
queries on (sparse) graphs with large diameters, such as road
networks, by representing a connected component with multiple spanning
trees to flatten them. We also want to make our approach
workload-aware, i.e., adapt it to a given ratio of queries and update
operations. Since our update operations are very efficient, we can
afford to add some overhead in the form of further optimizations when
faced with a high proportion of queries. Additionally, in the context
of workload-awareness we want to consider the distribution of
connectivity queries.  We also plan to investigate if our
  approach can be adapted to directed graphs.

\bibliographystyle{ACM-Reference-Format}
\interlinepenalty=10000


\begin{thebibliography}{51}


\ifx \showCODEN    \undefined \def \showCODEN     #1{\unskip}     \fi
\ifx \showDOI      \undefined \def \showDOI       #1{#1}\fi
\ifx \showISBNx    \undefined \def \showISBNx     #1{\unskip}     \fi
\ifx \showISBNxiii \undefined \def \showISBNxiii  #1{\unskip}     \fi
\ifx \showISSN     \undefined \def \showISSN      #1{\unskip}     \fi
\ifx \showLCCN     \undefined \def \showLCCN      #1{\unskip}     \fi
\ifx \shownote     \undefined \def \shownote      #1{#1}          \fi
\ifx \showarticletitle \undefined \def \showarticletitle #1{#1}   \fi
\ifx \showURL      \undefined \def \showURL       {\relax}        \fi
\providecommand\bibfield[2]{#2}
\providecommand\bibinfo[2]{#2}
\providecommand\natexlab[1]{#1}
\providecommand\showeprint[2][]{arXiv:#2}

\bibitem[\protect\citeauthoryear{??}{dat}{2021}]%
        {dataset_Stackoverflow}
 \bibinfo{year}{2021}\natexlab{}.
\newblock \bibinfo{title}{SNAP: Stack Overflow temporal network.}
\newblock
\newblock
\urldef\tempurl%
\url{http://snap.stanford.edu/data/sx-stackoverflow.html}
\showURL{%
Retrieved October 21, 2021 from \tempurl}


\bibitem[\protect\citeauthoryear{??}{KON}{2022}]%
        {KONECT_diameter}
 \bibinfo{year}{2022}\natexlab{}.
\newblock \bibinfo{title}{KONECT: The KONECT Project}.
\newblock
\newblock
\urldef\tempurl%
\url{http://konect.cc/statistics/diam/}
\showURL{%
Retrieved June 02, 2022 from \tempurl}


\bibitem[\protect\citeauthoryear{Alberts, Cattaneo, and Italiano}{Alberts
  et~al\mbox{.}}{1997}]%
        {David1997experiment}
\bibfield{author}{\bibinfo{person}{David Alberts}, \bibinfo{person}{Giuseppe
  Cattaneo}, {and} \bibinfo{person}{Giuseppe~F. Italiano}.}
  \bibinfo{year}{1997}\natexlab{}.
\newblock \showarticletitle{An Empirical Study of Dynamic Graph Algorithms}.
\newblock \bibinfo{journal}{\emph{ACM J. Exp. Algorithmics}}
  \bibinfo{volume}{2} (\bibinfo{date}{Jan.} \bibinfo{year}{1997}),
  \bibinfo{pages}{5–es}.
\newblock
\showISSN{1084-6654}
\urldef\tempurl%
\url{https://doi.org/10.1145/264216.264223}
\showDOI{\tempurl}


\bibitem[\protect\citeauthoryear{Ammar, Groeneveld, Bhagavatula, Beltagy,
  Crawford, Downey, Dunkelberger, Elgohary, Feldman, Ha, Kinney, Kohlmeier, Lo,
  Murray, Ooi, Peters, Power, Skjonsberg, Wang, Wilhelm, Yuan, van Zuylen, and
  Etzioni}{Ammar et~al\mbox{.}}{2018}]%
        {ammar}
\bibfield{author}{\bibinfo{person}{Waleed Ammar}, \bibinfo{person}{Dirk
  Groeneveld}, \bibinfo{person}{Chandra Bhagavatula}, \bibinfo{person}{Iz
  Beltagy}, \bibinfo{person}{Miles Crawford}, \bibinfo{person}{Doug Downey},
  \bibinfo{person}{Jason Dunkelberger}, \bibinfo{person}{Ahmed Elgohary},
  \bibinfo{person}{Sergey Feldman}, \bibinfo{person}{Vu Ha},
  \bibinfo{person}{Rodney Kinney}, \bibinfo{person}{Sebastian Kohlmeier},
  \bibinfo{person}{Kyle Lo}, \bibinfo{person}{Tyler Murray},
  \bibinfo{person}{Hsu-Han Ooi}, \bibinfo{person}{Matthew Peters},
  \bibinfo{person}{Joanna Power}, \bibinfo{person}{Sam Skjonsberg},
  \bibinfo{person}{Lucy~Lu Wang}, \bibinfo{person}{Chris Wilhelm},
  \bibinfo{person}{Zheng Yuan}, \bibinfo{person}{Madeleine van Zuylen}, {and}
  \bibinfo{person}{Oren Etzioni}.} \bibinfo{year}{2018}\natexlab{}.
\newblock \showarticletitle{Construction of the Literature Graph in Semantic
  Scholar}. In \bibinfo{booktitle}{\emph{NAACL}}.
\newblock
\urldef\tempurl%
\url{https://www.semanticscholar.org/paper/09e3cf5704bcb16e6657f6ceed70e93373a54618}
\showURL{%
\tempurl}


\bibitem[\protect\citeauthoryear{Bramandia, Choi, and Ng}{Bramandia
  et~al\mbox{.}}{2009}]%
        {bramandia2009incremental}
\bibfield{author}{\bibinfo{person}{Ramadhana Bramandia}, \bibinfo{person}{Byron
  Choi}, {and} \bibinfo{person}{Wee~Keong Ng}.}
  \bibinfo{year}{2009}\natexlab{}.
\newblock \showarticletitle{Incremental maintenance of 2-hop labeling of large
  graphs}.
\newblock \bibinfo{journal}{\emph{IEEE Transactions on Knowledge and Data
  Engineering}} \bibinfo{volume}{22}, \bibinfo{number}{5}
  (\bibinfo{year}{2009}), \bibinfo{pages}{682--698}.
\newblock


\bibitem[\protect\citeauthoryear{Cheng, Huang, Wu, and Fu}{Cheng
  et~al\mbox{.}}{2013}]%
        {TF_label}
\bibfield{author}{\bibinfo{person}{James Cheng}, \bibinfo{person}{Silu Huang},
  \bibinfo{person}{Huanhuan Wu}, {and} \bibinfo{person}{Ada Wai-Chee Fu}.}
  \bibinfo{year}{2013}\natexlab{}.
\newblock \showarticletitle{TF-Label: A Topological-Folding Labeling Scheme for
  Reachability Querying in a Large Graph} \emph{(\bibinfo{series}{SIGMOD
  '13})}. \bibinfo{publisher}{Association for Computing Machinery},
  \bibinfo{address}{New York, NY, USA}, 12.
\newblock
\showISBNx{9781450320375}
\urldef\tempurl%
\url{https://doi.org/10.1145/2463676.2465286}
\showDOI{\tempurl}


\bibitem[\protect\citeauthoryear{Chin and Houck}{Chin and Houck}{1978}]%
        {Chin78}
\bibfield{author}{\bibinfo{person}{Francis Chin} {and} \bibinfo{person}{David
  Houck}.} \bibinfo{year}{1978}\natexlab{}.
\newblock \showarticletitle{Algorithms for updating minimal spanning trees}.
\newblock \bibinfo{journal}{\emph{J. Comput. System Sci.}}
  \bibinfo{volume}{16}, \bibinfo{number}{3} (\bibinfo{year}{1978}),
  \bibinfo{pages}{333--344}.
\newblock
\showISSN{0022-0000}
\urldef\tempurl%
\url{https://doi.org/10.1016/0022-0000(78)90022-3}
\showDOI{\tempurl}


\bibitem[\protect\citeauthoryear{Cohen, Halperin, Kaplan, and Zwick}{Cohen
  et~al\mbox{.}}{2002}]%
        {Labeling_2hop}
\bibfield{author}{\bibinfo{person}{Edith Cohen}, \bibinfo{person}{Eran
  Halperin}, \bibinfo{person}{Haim Kaplan}, {and} \bibinfo{person}{Uri Zwick}.}
  \bibinfo{year}{2002}\natexlab{}.
\newblock \showarticletitle{Reachability and Distance Queries via 2-Hop
  Labels}. In \bibinfo{booktitle}{\emph{Proceedings of the Thirteenth Annual
  ACM-SIAM Symposium on Discrete Algorithms}} (San Francisco, California)
  \emph{(\bibinfo{series}{SODA ’02})}. \bibinfo{publisher}{SIAM},
  \bibinfo{address}{USA}, \bibinfo{pages}{937–946}.
\newblock
\showISBNx{089871513X}


\bibitem[\protect\citeauthoryear{Cormen, Leiserson, Rivest, and Stein}{Cormen
  et~al\mbox{.}}{2009}]%
        {cormen2009introduction}
\bibfield{author}{\bibinfo{person}{Thomas~H Cormen}, \bibinfo{person}{Charles~E
  Leiserson}, \bibinfo{person}{Ronald~L Rivest}, {and}
  \bibinfo{person}{Clifford Stein}.} \bibinfo{year}{2009}\natexlab{}.
\newblock \bibinfo{booktitle}{\emph{Introduction to algorithms}}.
\newblock \bibinfo{publisher}{MIT press}.
\newblock


\bibitem[\protect\citeauthoryear{Dobrynin, Entringer, and Gutman}{Dobrynin
  et~al\mbox{.}}{2001}]%
        {dobrynin2001wiener}
\bibfield{author}{\bibinfo{person}{Andrey~A Dobrynin}, \bibinfo{person}{Roger
  Entringer}, {and} \bibinfo{person}{Ivan Gutman}.}
  \bibinfo{year}{2001}\natexlab{}.
\newblock \showarticletitle{Wiener index of trees: theory and applications}.
\newblock \bibinfo{journal}{\emph{Acta Applicandae Mathematica}}
  \bibinfo{volume}{66}, \bibinfo{number}{3} (\bibinfo{year}{2001}),
  \bibinfo{pages}{211--249}.
\newblock


\bibitem[\protect\citeauthoryear{Doraiswamy and Natarajan}{Doraiswamy and
  Natarajan}{2009}]%
        {doraiswamy2009efficient}
\bibfield{author}{\bibinfo{person}{Harish Doraiswamy} {and}
  \bibinfo{person}{Vijay Natarajan}.} \bibinfo{year}{2009}\natexlab{}.
\newblock \showarticletitle{Efficient algorithms for computing Reeb graphs}.
\newblock \bibinfo{journal}{\emph{Computational Geometry}}
  \bibinfo{volume}{42}, \bibinfo{number}{6-7} (\bibinfo{year}{2009}),
  \bibinfo{pages}{606--616}.
\newblock


\bibitem[\protect\citeauthoryear{Eppstein}{Eppstein}{1992}]%
        {Eppstein92}
\bibfield{author}{\bibinfo{person}{D. Eppstein}.}
  \bibinfo{year}{1992}\natexlab{}.
\newblock \showarticletitle{Sparsification-a technique for speeding up dynamic
  graph algorithms}. In \bibinfo{booktitle}{\emph{Proc. of 33rd Annual
  Symposium on Foundations of Computer Science (FOCS'92)}}.
  \bibinfo{pages}{60--69}.
\newblock
\urldef\tempurl%
\url{https://doi.org/10.1109/SFCS.1992.267818}
\showDOI{\tempurl}


\bibitem[\protect\citeauthoryear{Eppstein, Galil, Italiano, and
  Nissenzweig}{Eppstein et~al\mbox{.}}{1997}]%
        {Eppstein97}
\bibfield{author}{\bibinfo{person}{David Eppstein}, \bibinfo{person}{Zvi
  Galil}, \bibinfo{person}{Giuseppe~F. Italiano}, {and} \bibinfo{person}{Amnon
  Nissenzweig}.} \bibinfo{year}{1997}\natexlab{}.
\newblock \showarticletitle{Sparsification—a Technique for Speeding up
  Dynamic Graph Algorithms}.
\newblock \bibinfo{journal}{\emph{J. ACM}} \bibinfo{volume}{44},
  \bibinfo{number}{5} (\bibinfo{date}{Sept.} \bibinfo{year}{1997}),
  \bibinfo{pages}{669–696}.
\newblock
\showISSN{0004-5411}
\urldef\tempurl%
\url{https://doi.org/10.1145/265910.265914}
\showDOI{\tempurl}


\bibitem[\protect\citeauthoryear{Eyal and Halperin}{Eyal and Halperin}{2005}]%
        {eyal2005improved}
\bibfield{author}{\bibinfo{person}{Eran Eyal} {and} \bibinfo{person}{Dan
  Halperin}.} \bibinfo{year}{2005}\natexlab{}.
\newblock \showarticletitle{Improved maintenance of molecular surfaces using
  dynamic graph connectivity}. In \bibinfo{booktitle}{\emph{International
  Workshop on Algorithms in Bioinformatics}}. Springer,
  \bibinfo{pages}{401--413}.
\newblock


\bibitem[\protect\citeauthoryear{Frederickson}{Frederickson}{1983}]%
        {Fred83}
\bibfield{author}{\bibinfo{person}{Greg~N. Frederickson}.}
  \bibinfo{year}{1983}\natexlab{}.
\newblock \showarticletitle{Data Structures for On-Line Updating of Minimum
  Spanning Trees}. In \bibinfo{booktitle}{\emph{Proc. of the 15th Annual ACM
  Symposium on Theory of Computing (STOC'83)}}. \bibinfo{publisher}{Association
  for Computing Machinery}, \bibinfo{address}{New York, NY, USA},
  \bibinfo{pages}{252–257}.
\newblock
\showISBNx{0897910990}
\urldef\tempurl%
\url{https://doi.org/10.1145/800061.808754}
\showDOI{\tempurl}


\bibitem[\protect\citeauthoryear{Gibbons}{Gibbons}{1985}]%
        {gibbons1985algorithmic}
\bibfield{author}{\bibinfo{person}{Alan Gibbons}.}
  \bibinfo{year}{1985}\natexlab{}.
\newblock \bibinfo{booktitle}{\emph{Algorithmic graph theory}}.
\newblock \bibinfo{publisher}{Cambridge university press}.
\newblock


\bibitem[\protect\citeauthoryear{Hegeman and Iosup}{Hegeman and Iosup}{2018}]%
        {Hegeman18}
\bibfield{author}{\bibinfo{person}{Tim Hegeman} {and}
  \bibinfo{person}{Alexandru Iosup}.} \bibinfo{year}{2018}\natexlab{}.
\newblock \showarticletitle{Survey of Graph Analysis Applications}.
\newblock \bibinfo{journal}{\emph{CoRR}}  \bibinfo{volume}{abs/1807.00382}
  (\bibinfo{year}{2018}).
\newblock
\showeprint[arXiv]{1807.00382}
\urldef\tempurl%
\url{http://arxiv.org/abs/1807.00382}
\showURL{%
\tempurl}


\bibitem[\protect\citeauthoryear{Helmer, Neumann, and Moerkotte}{Helmer
  et~al\mbox{.}}{2003}]%
        {Helmer03}
\bibfield{author}{\bibinfo{person}{Sven Helmer}, \bibinfo{person}{Thomas
  Neumann}, {and} \bibinfo{person}{Guido Moerkotte}.}
  \bibinfo{year}{2003}\natexlab{}.
\newblock \showarticletitle{A Robust Scheme for Multilevel Extendible Hashing}.
  In \bibinfo{booktitle}{\emph{Proc. 18th Int. Sym. on Computer and Information
  Sciences (ISCIS)}}. \bibinfo{address}{Antalya, Turkey},
  \bibinfo{pages}{220--227}.
\newblock


\bibitem[\protect\citeauthoryear{Henzinger and King}{Henzinger and
  King}{1995}]%
        {henzinger1995randomized}
\bibfield{author}{\bibinfo{person}{Monika~Rauch Henzinger} {and}
  \bibinfo{person}{Valerie King}.} \bibinfo{year}{1995}\natexlab{}.
\newblock \showarticletitle{Randomized dynamic graph algorithms with
  polylogarithmic time per operation}. In \bibinfo{booktitle}{\emph{Proc. of
  the 27th annual ACM symposium on Theory of computing (STOC'95)}}.
  \bibinfo{pages}{519--527}.
\newblock


\bibitem[\protect\citeauthoryear{Henzinger and King}{Henzinger and
  King}{1997}]%
        {henzinger97}
\bibfield{author}{\bibinfo{person}{Monika~Rauch Henzinger} {and}
  \bibinfo{person}{Valerie King}.} \bibinfo{year}{1997}\natexlab{}.
\newblock \showarticletitle{Maintaining Minimum Spanning Trees in Dynamic
  Graphs}. In \bibinfo{booktitle}{\emph{Proc. of 24th Int. Colloquium on
  Automata, Languages and Programming (ICALP'97)}}. \bibinfo{address}{Bologna,
  Italy}, \bibinfo{pages}{594--604}.
\newblock
\urldef\tempurl%
\url{https://doi.org/10.1007/3-540-63165-8\_214}
\showDOI{\tempurl}


\bibitem[\protect\citeauthoryear{Henzinger and King}{Henzinger and
  King}{1999}]%
        {henzinger99}
\bibfield{author}{\bibinfo{person}{Monika~Rauch Henzinger} {and}
  \bibinfo{person}{Valerie King}.} \bibinfo{year}{1999}\natexlab{}.
\newblock \showarticletitle{Randomized Fully Dynamic Graph Algorithms with
  Polylogarithmic Time per Operation}.
\newblock \bibinfo{journal}{\emph{J. {ACM}}} \bibinfo{volume}{46},
  \bibinfo{number}{4} (\bibinfo{year}{1999}), \bibinfo{pages}{502--516}.
\newblock
\urldef\tempurl%
\url{https://doi.org/10.1145/320211.320215}
\showDOI{\tempurl}


\bibitem[\protect\citeauthoryear{Henzinger and King}{Henzinger and
  King}{2001}]%
        {henzinger01}
\bibfield{author}{\bibinfo{person}{Monika~Rauch Henzinger} {and}
  \bibinfo{person}{Valerie King}.} \bibinfo{year}{2001}\natexlab{}.
\newblock \showarticletitle{Maintaining Minimum Spanning Forests in Dynamic
  Graphs}.
\newblock \bibinfo{journal}{\emph{{SIAM} J. Comput.}} \bibinfo{volume}{31},
  \bibinfo{number}{2} (\bibinfo{year}{2001}), \bibinfo{pages}{364--374}.
\newblock
\urldef\tempurl%
\url{https://doi.org/10.1137/S0097539797327209}
\showDOI{\tempurl}


\bibitem[\protect\citeauthoryear{Henzinger, King, and Warnow}{Henzinger
  et~al\mbox{.}}{1999}]%
        {henzinger1999constructing}
\bibfield{author}{\bibinfo{person}{Monika~Rauch Henzinger},
  \bibinfo{person}{Valerie King}, {and} \bibinfo{person}{Tandy Warnow}.}
  \bibinfo{year}{1999}\natexlab{}.
\newblock \showarticletitle{Constructing a tree from homeomorphic subtrees,
  with applications to computational evolutionary biology}.
\newblock \bibinfo{journal}{\emph{Algorithmica}} \bibinfo{volume}{24},
  \bibinfo{number}{1} (\bibinfo{year}{1999}), \bibinfo{pages}{1--13}.
\newblock


\bibitem[\protect\citeauthoryear{Holm, de~Lichtenberg, and Thorup}{Holm
  et~al\mbox{.}}{2001}]%
        {Holm2001}
\bibfield{author}{\bibinfo{person}{Jacob Holm}, \bibinfo{person}{Kristian de
  Lichtenberg}, {and} \bibinfo{person}{Mikkel Thorup}.}
  \bibinfo{year}{2001}\natexlab{}.
\newblock \showarticletitle{Poly-Logarithmic Deterministic Fully-Dynamic
  Algorithms for Connectivity, Minimum Spanning Tree, 2-Edge, and
  Biconnectivity}.
\newblock \bibinfo{journal}{\emph{J. ACM}} \bibinfo{volume}{48},
  \bibinfo{number}{4} (\bibinfo{date}{July} \bibinfo{year}{2001}),
  \bibinfo{pages}{723–760}.
\newblock
\showISSN{0004-5411}
\urldef\tempurl%
\url{https://doi.org/10.1145/502090.502095}
\showDOI{\tempurl}


\bibitem[\protect\citeauthoryear{Hopcroft and Tarjan}{Hopcroft and
  Tarjan}{1973}]%
        {Hop73}
\bibfield{author}{\bibinfo{person}{John Hopcroft} {and} \bibinfo{person}{Robert
  Tarjan}.} \bibinfo{year}{1973}\natexlab{}.
\newblock \showarticletitle{Algorithm 447: Efficient Algorithms for Graph
  Manipulation}.
\newblock \bibinfo{journal}{\emph{Commun. ACM}} \bibinfo{volume}{16},
  \bibinfo{number}{6} (\bibinfo{date}{June} \bibinfo{year}{1973}),
  \bibinfo{pages}{372–378}.
\newblock
\showISSN{0001-0782}
\urldef\tempurl%
\url{https://doi.org/10.1145/362248.362272}
\showDOI{\tempurl}


\bibitem[\protect\citeauthoryear{Huang, Huang, Kopelowitz, and Pettie}{Huang
  et~al\mbox{.}}{2017}]%
        {huang2017fully}
\bibfield{author}{\bibinfo{person}{Shang-En Huang}, \bibinfo{person}{Dawei
  Huang}, \bibinfo{person}{Tsvi Kopelowitz}, {and} \bibinfo{person}{Seth
  Pettie}.} \bibinfo{year}{2017}\natexlab{}.
\newblock \showarticletitle{Fully dynamic connectivity in O (log n (log log n)
  2) amortized expected time}. In \bibinfo{booktitle}{\emph{Proceedings of the
  twenty-eighth Annual ACM-SIAM Symposium on Discrete Algorithms}}. SIAM,
  \bibinfo{pages}{510--520}.
\newblock


\bibitem[\protect\citeauthoryear{Iyer, Karger, Rahul, and Thorup}{Iyer
  et~al\mbox{.}}{2002}]%
        {Raj2001experiment}
\bibfield{author}{\bibinfo{person}{Raj Iyer}, \bibinfo{person}{David Karger},
  \bibinfo{person}{Hariharan Rahul}, {and} \bibinfo{person}{Mikkel Thorup}.}
  \bibinfo{year}{2002}\natexlab{}.
\newblock \showarticletitle{An Experimental Study of Polylogarithmic, Fully
  Dynamic, Connectivity Algorithms}.
\newblock \bibinfo{journal}{\emph{ACM J. Exp. Algorithmics}}
  \bibinfo{volume}{6} (\bibinfo{date}{Dec.} \bibinfo{year}{2002}),
  \bibinfo{pages}{4–es}.
\newblock
\showISSN{1084-6654}
\urldef\tempurl%
\url{https://doi.org/10.1145/945394.945398}
\showDOI{\tempurl}


\bibitem[\protect\citeauthoryear{Jin, Xiang, Ruan, and Fuhry}{Jin
  et~al\mbox{.}}{2009}]%
        {3hop}
\bibfield{author}{\bibinfo{person}{Ruoming Jin}, \bibinfo{person}{Yang Xiang},
  \bibinfo{person}{Ning Ruan}, {and} \bibinfo{person}{David Fuhry}.}
  \bibinfo{year}{2009}\natexlab{}.
\newblock \showarticletitle{3-HOP: A High-Compression Indexing Scheme for
  Reachability Query}. In \bibinfo{booktitle}{\emph{Proceedings of the 2009 ACM
  SIGMOD International Conference on Management of Data}} (Providence, Rhode
  Island, USA) \emph{(\bibinfo{series}{SIGMOD '09})}.
  \bibinfo{publisher}{Association for Computing Machinery},
  \bibinfo{address}{New York, NY, USA}, \bibinfo{pages}{813–826}.
\newblock
\showISBNx{9781605585512}
\urldef\tempurl%
\url{https://doi.org/10.1145/1559845.1559930}
\showDOI{\tempurl}


\bibitem[\protect\citeauthoryear{Jordan}{Jordan}{1869}]%
        {Jordan1869}
\bibfield{author}{\bibinfo{person}{Camille Jordan}.}
  \bibinfo{year}{1869}\natexlab{}.
\newblock \showarticletitle{Sur les assemblages de lignes}.
\newblock \bibinfo{journal}{\emph{Journal f{\"u}r die reine und angewandte
  {Mathematik}}} \bibinfo{volume}{1869}, \bibinfo{number}{70}
  (\bibinfo{year}{1869}), \bibinfo{pages}{185--190}.
\newblock
\urldef\tempurl%
\url{https://doi.org/doi:10.1515/crll.1869.70.185}
\showDOI{\tempurl}


\bibitem[\protect\citeauthoryear{Kapron, King, and Mountjoy}{Kapron
  et~al\mbox{.}}{2013}]%
        {Kapron13}
\bibfield{author}{\bibinfo{person}{Bruce~M. Kapron}, \bibinfo{person}{Valerie
  King}, {and} \bibinfo{person}{Ben Mountjoy}.}
  \bibinfo{year}{2013}\natexlab{}.
\newblock \showarticletitle{Dynamic graph connectivity in polylogarithmic worst
  case time}. In \bibinfo{booktitle}{\emph{Proc. of the 24th Annual {ACM-SIAM}
  Symposium on Discrete Algorithms, {SODA'13}}}. \bibinfo{address}{New Orleans,
  Louisiana}, \bibinfo{pages}{1131--1142}.
\newblock
\urldef\tempurl%
\url{https://doi.org/10.1137/1.9781611973105.81}
\showDOI{\tempurl}


\bibitem[\protect\citeauthoryear{Kejlberg{-}Rasmussen, Kopelowitz, Pettie, and
  Thorup}{Kejlberg{-}Rasmussen et~al\mbox{.}}{2016}]%
        {Kejlberg16}
\bibfield{author}{\bibinfo{person}{Casper Kejlberg{-}Rasmussen},
  \bibinfo{person}{Tsvi Kopelowitz}, \bibinfo{person}{Seth Pettie}, {and}
  \bibinfo{person}{Mikkel Thorup}.} \bibinfo{year}{2016}\natexlab{}.
\newblock \showarticletitle{Faster Worst Case Deterministic Dynamic
  Connectivity}. In \bibinfo{booktitle}{\emph{24th Annual European Symposium on
  Algorithms (ESA'16)}}, \bibfield{editor}{\bibinfo{person}{Piotr Sankowski}
  {and} \bibinfo{person}{Christos~D. Zaroliagis}} (Eds.).
  \bibinfo{address}{Aarhus, Denmark}, \bibinfo{pages}{53:1--53:15}.
\newblock
\urldef\tempurl%
\url{https://doi.org/10.4230/LIPIcs.ESA.2016.53}
\showDOI{\tempurl}


\bibitem[\protect\citeauthoryear{Lyu, Li, He, and Gong}{Lyu
  et~al\mbox{.}}{2021}]%
        {DBL}
\bibfield{author}{\bibinfo{person}{Qiuyi Lyu}, \bibinfo{person}{Yuchen Li},
  \bibinfo{person}{Bingsheng He}, {and} \bibinfo{person}{Bin Gong}.}
  \bibinfo{year}{2021}\natexlab{}.
\newblock \showarticletitle{DBL: Efficient Reachability Queries on Dynamic
  Graphs}. In \bibinfo{booktitle}{\emph{International Conference on Database
  Systems for Advanced Applications}}. Springer, \bibinfo{pages}{761--777}.
\newblock


\bibitem[\protect\citeauthoryear{Mislove}{Mislove}{2009}]%
        {mislove09}
\bibfield{author}{\bibinfo{person}{Alan Mislove}.}
  \bibinfo{year}{2009}\natexlab{}.
\newblock \emph{\bibinfo{title}{Online Social Networks: Measurement, Analysis,
  and Applications to Distributed Information Systems}}.
\newblock \bibinfo{thesistype}{Ph.D. Dissertation}. \bibinfo{school}{Rice
  University, Department of Computer Science}.
\newblock


\bibitem[\protect\citeauthoryear{Rossi and Ahmed}{Rossi and Ahmed}{2015}]%
        {nr}
\bibfield{author}{\bibinfo{person}{Ryan~A. Rossi} {and}
  \bibinfo{person}{Nesreen~K. Ahmed}.} \bibinfo{year}{2015}\natexlab{}.
\newblock \showarticletitle{The Network Data Repository with Interactive Graph
  Analytics and Visualization}. In \bibinfo{booktitle}{\emph{AAAI}}.
\newblock
\urldef\tempurl%
\url{http://networkrepository.com}
\showURL{%
\tempurl}


\bibitem[\protect\citeauthoryear{Sahu, Mhedhbi, Salihoglu, Lin, and
  \"{O}zsu}{Sahu et~al\mbox{.}}{2017}]%
        {vldb_largegraphs}
\bibfield{author}{\bibinfo{person}{Siddhartha Sahu}, \bibinfo{person}{Amine
  Mhedhbi}, \bibinfo{person}{Semih Salihoglu}, \bibinfo{person}{Jimmy Lin},
  {and} \bibinfo{person}{M.~Tamer \"{O}zsu}.} \bibinfo{year}{2017}\natexlab{}.
\newblock \showarticletitle{The Ubiquity of Large Graphs and Surprising
  Challenges of Graph Processing}.
\newblock \bibinfo{journal}{\emph{Proc. VLDB Endow.}} \bibinfo{volume}{11},
  \bibinfo{number}{4} (\bibinfo{date}{Dec.} \bibinfo{year}{2017}),
  \bibinfo{pages}{420–431}.
\newblock
\showISSN{2150-8097}
\urldef\tempurl%
\url{https://doi.org/10.1145/3186728.3164139}
\showDOI{\tempurl}


\bibitem[\protect\citeauthoryear{Sakr, Bonifati, Voigt, Iosup, Ammar, Angles,
  Aref, Arenas, Besta, Boncz, Daudjee, Valle, Dumbrava, Hartig, Haslhofer,
  Hegeman, Hidders, Hose, Iamnitchi, Kalavri, Kapp, Martens, \"{O}zsu, Peukert,
  Plantikow, Ragab, Ripeanu, Salihoglu, Schulz, Selmer, Sequeda, Shinavier,
  Sz\'{a}rnyas, Tommasini, Tumeo, Uta, Varbanescu, Wu, Yakovets, Yan, and
  Yoneki}{Sakr et~al\mbox{.}}{2021}]%
        {Sakr21}
\bibfield{author}{\bibinfo{person}{Sherif Sakr}, \bibinfo{person}{Angela
  Bonifati}, \bibinfo{person}{Hannes Voigt}, \bibinfo{person}{Alexandru Iosup},
  \bibinfo{person}{Khaled Ammar}, \bibinfo{person}{Renzo Angles},
  \bibinfo{person}{Walid Aref}, \bibinfo{person}{Marcelo Arenas},
  \bibinfo{person}{Maciej Besta}, \bibinfo{person}{Peter~A. Boncz},
  \bibinfo{person}{Khuzaima Daudjee}, \bibinfo{person}{Emanuele~Della Valle},
  \bibinfo{person}{Stefania Dumbrava}, \bibinfo{person}{Olaf Hartig},
  \bibinfo{person}{Bernhard Haslhofer}, \bibinfo{person}{Tim Hegeman},
  \bibinfo{person}{Jan Hidders}, \bibinfo{person}{Katja Hose},
  \bibinfo{person}{Adriana Iamnitchi}, \bibinfo{person}{Vasiliki Kalavri},
  \bibinfo{person}{Hugo Kapp}, \bibinfo{person}{Wim Martens},
  \bibinfo{person}{M.~Tamer \"{O}zsu}, \bibinfo{person}{Eric Peukert},
  \bibinfo{person}{Stefan Plantikow}, \bibinfo{person}{Mohamed Ragab},
  \bibinfo{person}{Matei~R. Ripeanu}, \bibinfo{person}{Semih Salihoglu},
  \bibinfo{person}{Christian Schulz}, \bibinfo{person}{Petra Selmer},
  \bibinfo{person}{Juan~F. Sequeda}, \bibinfo{person}{Joshua Shinavier},
  \bibinfo{person}{G\'{a}bor Sz\'{a}rnyas}, \bibinfo{person}{Riccardo
  Tommasini}, \bibinfo{person}{Antonino Tumeo}, \bibinfo{person}{Alexandru
  Uta}, \bibinfo{person}{Ana~Lucia Varbanescu}, \bibinfo{person}{Hsiang-Yun
  Wu}, \bibinfo{person}{Nikolay Yakovets}, \bibinfo{person}{Da Yan}, {and}
  \bibinfo{person}{Eiko Yoneki}.} \bibinfo{year}{2021}\natexlab{}.
\newblock \showarticletitle{The Future is Big Graphs: A Community View on Graph
  Processing Systems}.
\newblock \bibinfo{journal}{\emph{Commun. ACM}} \bibinfo{volume}{64},
  \bibinfo{number}{9} (\bibinfo{date}{Aug.} \bibinfo{year}{2021}),
  \bibinfo{pages}{62--71}.
\newblock
\showISSN{0001-0782}
\urldef\tempurl%
\url{https://doi.org/10.1145/3434642}
\showDOI{\tempurl}


\bibitem[\protect\citeauthoryear{Seidel and Aragon}{Seidel and Aragon}{1996}]%
        {seidel1996randomized}
\bibfield{author}{\bibinfo{person}{Raimund Seidel} {and}
  \bibinfo{person}{Cecilia~R Aragon}.} \bibinfo{year}{1996}\natexlab{}.
\newblock \showarticletitle{Randomized search trees}.
\newblock \bibinfo{journal}{\emph{Algorithmica}} \bibinfo{volume}{16},
  \bibinfo{number}{4} (\bibinfo{year}{1996}), \bibinfo{pages}{464--497}.
\newblock


\bibitem[\protect\citeauthoryear{Shiloach and Even}{Shiloach and Even}{1981}]%
        {Shiloach81}
\bibfield{author}{\bibinfo{person}{Yossi Shiloach} {and}
  \bibinfo{person}{Shimon Even}.} \bibinfo{year}{1981}\natexlab{}.
\newblock \showarticletitle{An On-Line Edge-Deletion Problem}.
\newblock \bibinfo{journal}{\emph{J. ACM}} \bibinfo{volume}{28},
  \bibinfo{number}{1} (\bibinfo{date}{Jan.} \bibinfo{year}{1981}),
  \bibinfo{pages}{1–4}.
\newblock
\showISSN{0004-5411}
\urldef\tempurl%
\url{https://doi.org/10.1145/322234.322235}
\showDOI{\tempurl}


\bibitem[\protect\citeauthoryear{Spira and Pan}{Spira and Pan}{1975}]%
        {Spira75}
\bibfield{author}{\bibinfo{person}{P.M. Spira} {and} \bibinfo{person}{A. Pan}.}
  \bibinfo{year}{1975}\natexlab{}.
\newblock \showarticletitle{On Finding and Updating Spanning Trees and Shortest
  Paths}.
\newblock \bibinfo{journal}{\emph{SIAM J. Comput.}} \bibinfo{volume}{4},
  \bibinfo{number}{3} (\bibinfo{year}{1975}), \bibinfo{pages}{375--380}.
\newblock
\urldef\tempurl%
\url{https://doi.org/10.1137/0204032}
\showDOI{\tempurl}


\bibitem[\protect\citeauthoryear{Szpankowski}{Szpankowski}{1990}]%
        {Szpan90}
\bibfield{author}{\bibinfo{person}{Wojciech Szpankowski}.}
  \bibinfo{year}{1990}\natexlab{}.
\newblock \showarticletitle{Patricia Tries Again Revisited}.
\newblock \bibinfo{journal}{\emph{J. ACM}} \bibinfo{volume}{37},
  \bibinfo{number}{4} (\bibinfo{date}{Oct.} \bibinfo{year}{1990}),
  \bibinfo{pages}{691–711}.
\newblock
\showISSN{0004-5411}
\urldef\tempurl%
\url{https://doi.org/10.1145/96559.214080}
\showDOI{\tempurl}


\bibitem[\protect\citeauthoryear{Tarjan}{Tarjan}{1975}]%
        {Tarjan75}
\bibfield{author}{\bibinfo{person}{Robert~Endre Tarjan}.}
  \bibinfo{year}{1975}\natexlab{}.
\newblock \showarticletitle{Efficiency of a Good But Not Linear Set Union
  Algorithm}.
\newblock \bibinfo{journal}{\emph{J. ACM}} \bibinfo{volume}{22},
  \bibinfo{number}{2} (\bibinfo{date}{April} \bibinfo{year}{1975}),
  \bibinfo{pages}{215–225}.
\newblock
\showISSN{0004-5411}
\urldef\tempurl%
\url{https://doi.org/10.1145/321879.321884}
\showDOI{\tempurl}


\bibitem[\protect\citeauthoryear{Tarjan}{Tarjan}{1983}]%
        {tarjan1983data}
\bibfield{author}{\bibinfo{person}{Robert~Endre Tarjan}.}
  \bibinfo{year}{1983}\natexlab{}.
\newblock \bibinfo{booktitle}{\emph{Data structures and network algorithms}}.
\newblock \bibinfo{publisher}{SIAM}.
\newblock


\bibitem[\protect\citeauthoryear{Tarjan and Vishkin}{Tarjan and
  Vishkin}{1984}]%
        {tarjan1984finding}
\bibfield{author}{\bibinfo{person}{Robert~Endre Tarjan} {and}
  \bibinfo{person}{Uzi Vishkin}.} \bibinfo{year}{1984}\natexlab{}.
\newblock \showarticletitle{Finding biconnected componemts and computing tree
  functions in logarithmic parallel time}. In \bibinfo{booktitle}{\emph{25th
  Annual Symposium on Foundations of Computer Science, 1984.}} IEEE,
  \bibinfo{pages}{12--20}.
\newblock


\bibitem[\protect\citeauthoryear{Tarjan and Vishkin}{Tarjan and
  Vishkin}{1985}]%
        {tarjan1985efficient}
\bibfield{author}{\bibinfo{person}{Robert~E Tarjan} {and} \bibinfo{person}{Uzi
  Vishkin}.} \bibinfo{year}{1985}\natexlab{}.
\newblock \showarticletitle{An efficient parallel biconnectivity algorithm}.
\newblock \bibinfo{journal}{\emph{SIAM J. Comput.}} \bibinfo{volume}{14},
  \bibinfo{number}{4} (\bibinfo{year}{1985}), \bibinfo{pages}{862--874}.
\newblock


\bibitem[\protect\citeauthoryear{Thorup}{Thorup}{2000}]%
        {Thorup2000}
\bibfield{author}{\bibinfo{person}{Mikkel Thorup}.}
  \bibinfo{year}{2000}\natexlab{}.
\newblock \showarticletitle{Near-Optimal Fully-Dynamic Graph Connectivity}. In
  \bibinfo{booktitle}{\emph{Proceedings of the Thirty-Second Annual ACM
  Symposium on Theory of Computing}} (Portland, Oregon, USA)
  \emph{(\bibinfo{series}{STOC '00})}. \bibinfo{publisher}{Association for
  Computing Machinery}, \bibinfo{address}{New York, NY, USA},
  \bibinfo{pages}{343–350}.
\newblock
\showISBNx{1581131844}
\urldef\tempurl%
\url{https://doi.org/10.1145/335305.335345}
\showDOI{\tempurl}


\bibitem[\protect\citeauthoryear{Wei, Yu, Lu, and Jin}{Wei
  et~al\mbox{.}}{2018}]%
        {IP_label}
\bibfield{author}{\bibinfo{person}{Hao Wei}, \bibinfo{person}{Jeffrey~Xu Yu},
  \bibinfo{person}{Can Lu}, {and} \bibinfo{person}{Ruoming Jin}.}
  \bibinfo{year}{2018}\natexlab{}.
\newblock \showarticletitle{Reachability Querying: An Independent Permutation
  Labeling Approach}.
\newblock \bibinfo{journal}{\emph{The VLDB Journal}} \bibinfo{volume}{27},
  \bibinfo{number}{1} (\bibinfo{date}{Feb.} \bibinfo{year}{2018}),
  \bibinfo{pages}{1–26}.
\newblock
\showISSN{1066-8888}
\urldef\tempurl%
\url{https://doi.org/10.1007/s00778-017-0468-3}
\showDOI{\tempurl}


\bibitem[\protect\citeauthoryear{West et~al\mbox{.}}{West
  et~al\mbox{.}}{2001}]%
        {west2001introduction}
\bibfield{author}{\bibinfo{person}{Douglas~Brent West} {et~al\mbox{.}}}
  \bibinfo{year}{2001}\natexlab{}.
\newblock \bibinfo{booktitle}{\emph{Introduction to graph theory}}.
  Vol.~\bibinfo{volume}{2}.
\newblock \bibinfo{publisher}{Prentice hall Upper Saddle River}.
\newblock


\bibitem[\protect\citeauthoryear{Wulff-Nilsen}{Wulff-Nilsen}{2013}]%
        {wulff2013faster}
\bibfield{author}{\bibinfo{person}{Christian Wulff-Nilsen}.}
  \bibinfo{year}{2013}\natexlab{}.
\newblock \showarticletitle{Faster deterministic fully-dynamic graph
  connectivity}. In \bibinfo{booktitle}{\emph{Proceedings of the twenty-fourth
  Annual ACM-SIAM Symposium on Discrete Algorithms}}. SIAM,
  \bibinfo{pages}{1757--1769}.
\newblock


\bibitem[\protect\citeauthoryear{Zaroliagis}{Zaroliagis}{2002}]%
        {Zaroliagis02}
\bibfield{author}{\bibinfo{person}{Christos~D. Zaroliagis}.}
  \bibinfo{year}{2002}\natexlab{}.
\newblock \bibinfo{booktitle}{\emph{Implementations and Experimental Studies of
  Dynamic Graph Algorithms}}.
\newblock \bibinfo{publisher}{Springer-Verlag}, \bibinfo{address}{Berlin,
  Heidelberg}, \bibinfo{pages}{2290–278}.
\newblock
\showISBNx{3540003460}


\bibitem[\protect\citeauthoryear{Zelinka}{Zelinka}{1968}]%
        {Zelinka68}
\bibfield{author}{\bibinfo{person}{Bohdan Zelinka}.}
  \bibinfo{year}{1968}\natexlab{}.
\newblock \showarticletitle{Medians and Peripherians of Trees}.
\newblock \bibinfo{journal}{\emph{Archivum Mathematicum}} \bibinfo{volume}{4},
  \bibinfo{number}{2} (\bibinfo{year}{1968}), \bibinfo{pages}{87--95}.
\newblock


\bibitem[\protect\citeauthoryear{Zhu, Lin, Wang, and Xiao}{Zhu
  et~al\mbox{.}}{2014}]%
        {TOL}
\bibfield{author}{\bibinfo{person}{Andy~Diwen Zhu}, \bibinfo{person}{Wenqing
  Lin}, \bibinfo{person}{Sibo Wang}, {and} \bibinfo{person}{Xiaokui Xiao}.}
  \bibinfo{year}{2014}\natexlab{}.
\newblock \showarticletitle{Reachability Queries on Large Dynamic Graphs: A
  Total Order Approach}. In \bibinfo{booktitle}{\emph{Proceedings of the 2014
  ACM SIGMOD International Conference on Management of Data}} (Snowbird, Utah,
  USA) \emph{(\bibinfo{series}{SIGMOD ’14})}. \bibinfo{publisher}{ACM},
  \bibinfo{address}{New York, NY, USA}, \bibinfo{pages}{1323–1334}.
\newblock
\showISBNx{9781450323765}
\urldef\tempurl%
\url{https://doi.org/10.1145/2588555.2612181}
\showDOI{\tempurl}


\end{thebibliography}

\inreport{
\section{Appendix}
\appendix

\section{Examples}

\subsection{BFS tree}
\label{exam_spanning_trees}

\begin{example}
  In Figure~\ref{fig:spanning_trees}, spanning trees $T_3$ and $T_4$
  are BFS trees with root $n_1$ and $n_2$, respectively.
\end{example}

\begin{figure}[htb]\centering
  \tikzstyle{every node}=[fill=black,circle,inner sep=0pt,minimum size=2.5pt,]
  \begin{subfigure}[b]{0.3\columnwidth}\centering
    \scalebox{0.8}{
    \begin{tikzpicture}
      \tikzstyle{every node}=[fill=black,circle,inner sep=0pt,minimum size=2.5pt,]
      \node [fill=red, label=above:$\mathsf{n}_{1}$] (1) at (-4.0, 0.0) {};
      \node [label=left:$\mathsf{n}_{2}$] (2) at (-5.0, -1.0) {};
      \node [label=left:$\mathsf{n}_{3}$] (3) at (-4.0, -1.0) {};
      \node [label=right:$\mathsf{n}_{4}$] (4) at (-3.0, -1.0) {};
      \node [label=below:$\mathsf{n}_{5}$] (5) at (-4.5, -2.0) {};
      \node [label=below:$\mathsf{n}_{6}$] (6) at (-3.5, -2.0) {};
      
      \draw[-] (1) to (2);
      \draw[-] (1) to (3);
      \draw[-] (1) to (4);
      \draw[-] (2) to (5);
      \draw[-] (4) to (6);
    \end{tikzpicture}
    }
    \caption{$T_3$}
  \end{subfigure}
  \quad
  \begin{subfigure}[b]{0.3\columnwidth}\centering
    \scalebox{0.8}{
    \begin{tikzpicture}
      \tikzstyle{every node}=[fill=black,circle,inner sep=0pt,minimum size=2.5pt,]
      \node [label=above:$\mathsf{n}_{1}$] (1) at (-4.0, 0.0) {};
      \node [fill=red, label=left:$\mathsf{n}_{2}$] (2) at (-5.0, -1.0) {};
      \node [label=left:$\mathsf{n}_{3}$] (3) at (-4.0, -1.0) {};
      \node [label=right:$\mathsf{n}_{4}$] (4) at (-3.0, -1.0) {};
      \node [label=below:$\mathsf{n}_{5}$] (5) at (-4.5, -2.0) {};
      \node [label=below:$\mathsf{n}_{6}$] (6) at (-3.5, -2.0) {};
      
      \draw[-] (1) to (2);
      \draw[-] (1) to (4);
      \draw[-] (3) to (5);
      \draw[-] (2) to (5);
      \draw[-] (4) to (6);
    \end{tikzpicture}
    }
    \caption{$T_4$}
  \end{subfigure}
  \quad
  \begin{subfigure}[b]{0.3\columnwidth}\centering
    \scalebox{0.8}{
    \begin{tikzpicture}
      \tikzstyle{every node}=[fill=black,circle,inner sep=0pt,minimum size=2.5pt,]
      \node [label=above:$\mathsf{n}_{1}$] (1) at (-4.0, 0.0) {};
      \node [label=left:$\mathsf{n}_{2}$] (2) at (-5.0, -1.0) {};
      \node [fill=red, label=left:$\mathsf{n}_{3}$] (3) at (-4.0, -1.0) {};
      \node [label=right:$\mathsf{n}_{4}$] (4) at (-3.0, -1.0) {};
      \node [label=below:$\mathsf{n}_{5}$] (5) at (-4.5, -2.0) {};
      \node [label=below:$\mathsf{n}_{6}$] (6) at (-3.5, -2.0) {};
      
      \draw[-] (1) to (3);
      \draw[-] (1) to (4);
      \draw[-] (3) to (5);
      \draw[-] (2) to (5);
      \draw[-] (4) to (6);
    \end{tikzpicture}
    }
    \caption{$T_5$}
  \end{subfigure}

  \caption{Different spanning trees for $C_1$ in
    Figure~\ref{fig:example_graph_G1}. The red nodes are the roots of
    the spanning trees.  $T_1$ in
    Figure~\ref{fig:graph_spanning_forest}, $T_3$ and $T_4$ are BFS
    trees for $C_1$.  $T_5$ is not a BFS tree for $C_1$.}
  \label{fig:spanning_trees}
\end{figure}
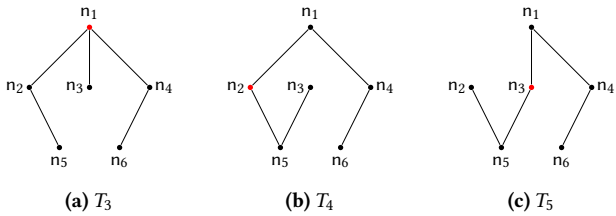

\section{Algorithms}

\subsection{Link}
\label{link_alg}
Algorithm~\ref{alg:link} shows the pseudocode of link operation.

\begin{algorithm2e}[htb]
  \small
  \caption{link($n_u$, $r_u$, $n_v$)}
  \label{alg:link}
  \SetKwInOut{Input}{input}
  \SetKwInOut{Output}{output}
  
  \Input{a node $n_u$ in D-tree $D$ with the root $r_u$, 
    the root $n_v$ of a D-tree
    currently not connected to $D$ via a tree edge}
  \Output{merged D-tree with new tree edge ($n_u, n_v$)}

  add $n_v$ to $n_u.children$ \\
  $n_v.parent = n_u$ \\

  $m = Null$ \tcp*[l]{new centroid}
  $i = n_u$ \\
  \While{$i \neq Null$} {

    $i.size = i.size + n_v.size$ \label{alg:link_increment}\\
    \lIf{$i.size$> $\sfrac{(r_u.size + n_v.size)}{2}$ and $m == Null$
    }{ \label{alg:link_newroot}
      $m = i$
    }
    
    $i = i.parent$ \\
  }

  \lIf{$m\neq Null$ and $m \neq r_u$} {
    $r_u$ = reroot($m$)
  }
  \Return $r_u$

\end{algorithm2e}

\subsection{Unlink}
\label{unlink_alg}
Algorithm~\ref{alg:unlink} shows the pseudocode of unlink operation.

\begin{algorithm2e}[htb]
  \small
  \caption{unlink($n_v$)}
  \label{alg:unlink}
  \SetKwInOut{Input}{input}
  \SetKwInOut{Output}{output}
  
  \Input{a non-root node $n_v$ in D-tree $D$}
  \Output{two D-trees, not connected via tree edges}

  $i = n_v$ \\
  \While{$i.parent \neq Null$} {
    $i = i.parent$ \\
    $i.size = i.size - n_v.size$ 
  }
  remove $n_v$ from $n_v.parent.children$ \\
  $n_v.parent = Null$ \\
  \Return ($n_v$, $i$)
\end{algorithm2e}

\subsection{Delete non-tree edge}

Algorithm~\ref{alg:deletente}
shows the pseudocode for the deletion of a non-tree edge.

\begin{algorithm2e}[htb]
  \small
  \caption{delete$_{nte}$($n_u$, $n_v$)}
  \label{alg:deletente}
  \SetKwInOut{Input}{input}
  \SetKwInOut{Output}{output}

  \Input{Tree nodes $n_u$ and $n_v$}
  \Output{Updated nodes $n_u$ and $n_v$}

  remove $n_u$ from $n_v.nte$ \\
  remove $n_v$ from $n_u.nte$ \\
\end{algorithm2e}

\section{Proofs}

\subsection{Proof for Theorem \ref{theorem:average_cost}}
\label{average_cost_proof}

\begin{proof}
When answering connectivity queries
$conn(u,v)$, we traverse from $u$ and $v$ to the roots $r_u$ and $r_v$
containing them ($r_u$ and $r_v$ can be equal). Let $d_R(u) = d_T(r_u,u)$ be
the distance between a node and the root of its spanning tree. The cost of
the traversal from node to root is directly proportional to
$d_R(u)$. Consequently, the total cost $c_{tot}$ of connectivity queries over
all pairs of nodes $u$ and $v$ ($u \neq v$) is equal to
$\sum_{u<v}(d_R(u) + d_R(v))$

\begin{align*}
  &= \frac{1}{2} \left( \sum_{u \in V} \sum_{v \in V} (d_R(u) + d_R(v)) 
  - \sum_{u = v \in V} (d_R(u) + d_R(v))   \right)\\
  &=  \frac{1}{2} \left( \sum_{u \in V} \sum_{v \in V} (d_R(u) + d_R(v)) \right)
  - \frac{1}{2} \cdot 2 \sum_{u \in V} d_R(u)   \\
  &=  \frac{1}{2} \left( \sum_{u \in V} \sum_{v \in V} d_R(u) +
  \sum_{u \in V} \sum_{v \in V} d_R(v) \right)
  - \sum_{u \in V} d_R(u)   \\
  &=  \frac{1}{2} \left( |V| \sum_{u \in V} d_R(u) +
  |V| \sum_{v \in V} d_R(v) \right)
  - \sum_{u \in V} d_R(u)   \\
  & = |V| \sum_{u \in V} d_R(u) - \sum_{u \in V} d_R(u) =
  (|V| - 1) \sum_{u \in V} d_R(u) \\
\end{align*}

\noindent
So, the average cost $c_{\mbox{\scriptsize{\emph{avg}}}}$ per query (assuming uniformly distributed
queries) is equal to 
\begin{align*}
 & \frac{\sum_{u<v}(d_R(u) + d_R(v))}{\binom{|V|}{2}} =
  \frac{(|V| - 1) \sum_{u \in V} d_R(u)}{\frac{|V| (|V| - 1)}{2}} = \\
 & \frac{2 \sum_{u \in V} d_R(u)}{|V|} = 2 E(d_R(u)) \\
\end{align*}
\end{proof}

\subsection{Proof for Lemma \ref{lemma:bfs_tree}}
\label{bfs_tree_proof}

\begin{proof}
  Let us assume that we have a BFS-tree $T$ with root $r$ in which the
  sum of distances between $r$ and all other nodes is not
  minimal. Thus, there is at least one node for which we can find a
  shorter path to the root: we call this node $u$. The current path
  from $r$ to $u$ in $T$ is $r, u_1^c, u_2^c, \dots, u_k^c, u$, while
  the shortest path is $r, u_1^s, u_2^s, \dots, u_l^s, u$ with $k>l$.
  This is a contradiction to the definition of a BFS-tree. As a
  BFS-tree expands level by level, the node $u$ would have already
  been reached after $l$ steps via the path containing the nodes
  $u_i^s$.
\end{proof}

\subsection{Proof for Lemma \ref{lemma:sc_sd}}
\label{lemma_sc_sd_proof}

  \begin{proof}
    Let $D$ be the maximal depth of $T$ and
    $V_d = \{ v | v \in V', d_T(r,v) = d \}$ the vertices in $T$ with depth
    $d$. By Definition~\ref{def:sumDist},

    \begin{align*}
   S_d(T) &= \sum_{v \in V'} d_T(r, v) \\
    &= \sum_{d = 1}^{D} d * |V_d| \\
    &= |V_1| + 2 * |V_2| + 3 * |V_3| + ... +  D * |V_D|\\
    &= |V_1| + |V_2| + |V_3| + ... + |V_D|\\
    &+ |V_2| + |V_3| + ... + |V_D|\\
    &+ |V_3| + ... + |V_D|\\
    & \dots\\
    &+ |V_{D-1}| + |V_D|\\
    &+ |V_D|\\
    &= \sum_{v \in V_1} size(v) + \sum_{v \in V_2} size(v) + \dots 
    + \sum_{v \in V_D} size(v)\\
    &= \sum_{d=1}^{D} \sum_{v \in V_d} size(v) 
    = \sum_{v \in V' \setminus r } size(v) = S_c(T)
  \end{align*}
    
  \end{proof}

\subsection{Proof for Lemma \ref{lemma:bfscentroid}}
\label{lemma_bfscentroid_proof}
\begin{proof}
According to Lemma~\ref{lemma:bfs_tree}, there is no other tree rooted at $r$
with a smaller value for $S_d$. Assume that $r$ is not a centroid of $T_m$.
Let $c_1, c_2, \dots, c_k$ be the children of $r$. Since $r$ is not a
centroid, one of the children of $r$, $c_j$, has a size greater than
$\sfrac{|V_m|}{2}$ (see Theorem~\ref{theo:centroid}). We designate $c_j$ as the
new root of $T_m$, by making $r$ a child of $c_j$, creating the tree
$T_{c_j}$. This pulls up $c_j$ and all its descendants by one level, while
pushing down $r$ and all its other children ($c_i \not= c_j$) by one level:

\begin{align*}
  S_d(T_{c_j}) &= S_d(T_m) - size(c_j) + 1 + \sum_{i \not= j} size(c_i) \\
  &= S_d(T_m) - size(c_j) + (size(r) - size(c_j)) \\
  &= S_d(T_m) + size(r) - 2 \cdot size(c_j) < S_d(T_m) 
\end{align*}

This contradicts that $S_d(T_m)$ is minimal.
\end{proof}

\section{Experiments}

\subsection{Distributions of distances between nodes and roots}
Figure~\ref{fig:distributions} gives a more detailed insight 
into the distribution of node depths in the various trees. We accumulate
the frequency of each node depth in spanning tress at all testing 
points, and calculate
the average frequency of each node depth. On average, 
the nodes in our D-trees are much closer to the roots. 
For small graphs (upper row of Figure~\ref{fig:distributions}), 
we are very close to opt. For large graphs (lower row of Figure~\ref{fig:distributions}), D-trees also outperform the other methods.

\begin{figure*}[htb]\centering
  \begin{subfigure}[c]{\textwidth}\centering
    \includegraphics{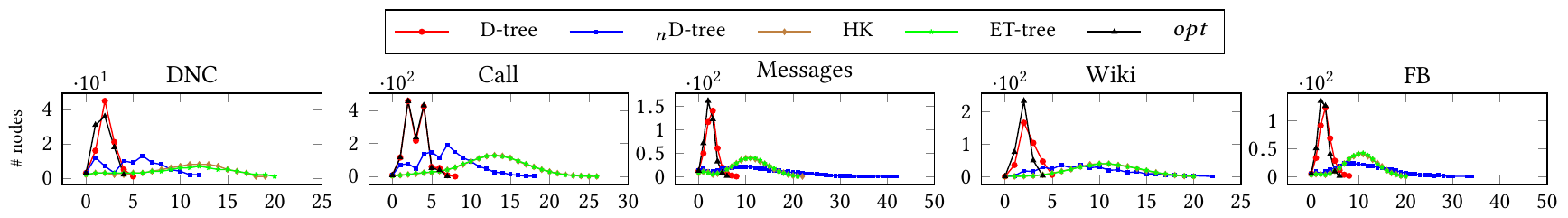}
  \end{subfigure}

  \begin{subfigure}[c]{\textwidth}\centering
    \includegraphics{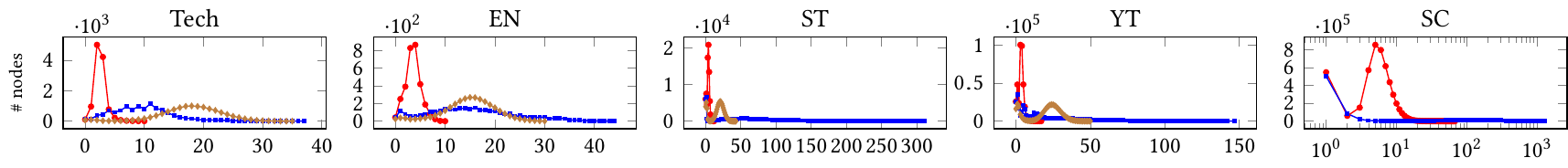}
  \end{subfigure}
  \caption{Distributions of average node depths. }
  \label{fig:distributions}
\end{figure*}

\subsection{Numbers of vertices and edges in graphs at each snapshot}

Figure~\ref{fig:num_vertices} and Figure~\ref{fig:num_edges} show the 
numbers of vertices and edges at each snapshot of the graphs respectively. 
In general, between two neighboring
snapshots, more (fewer) insertions of edges
than deletions of edges increases (decreases) the numbers of vertices and edges 
in the graphs. 

\begin{figure*}[htb] \centering
  \includegraphics{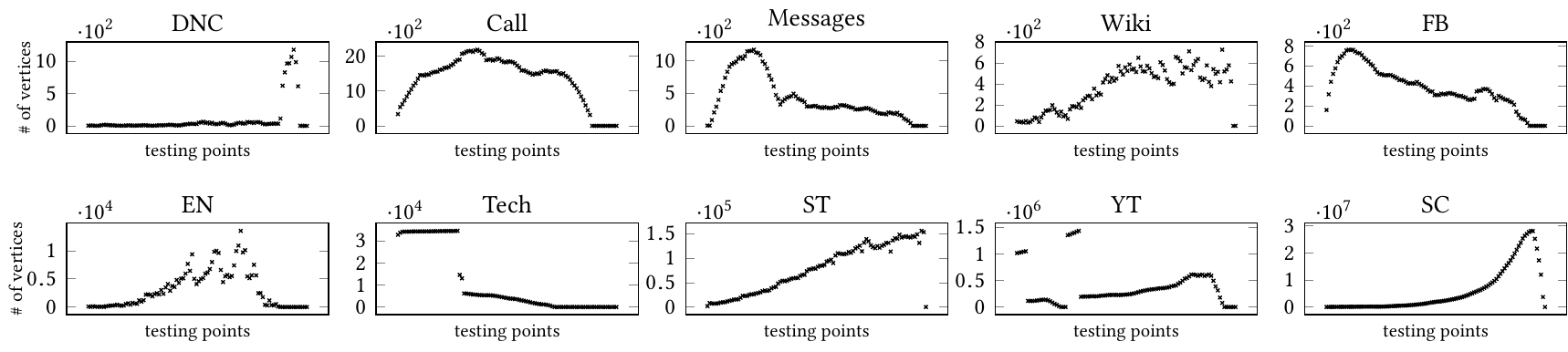}
  \caption{\# vertices in graphs.}
  \label{fig:num_vertices}
\end{figure*}

\begin{figure*}[htb] \centering
  \includegraphics{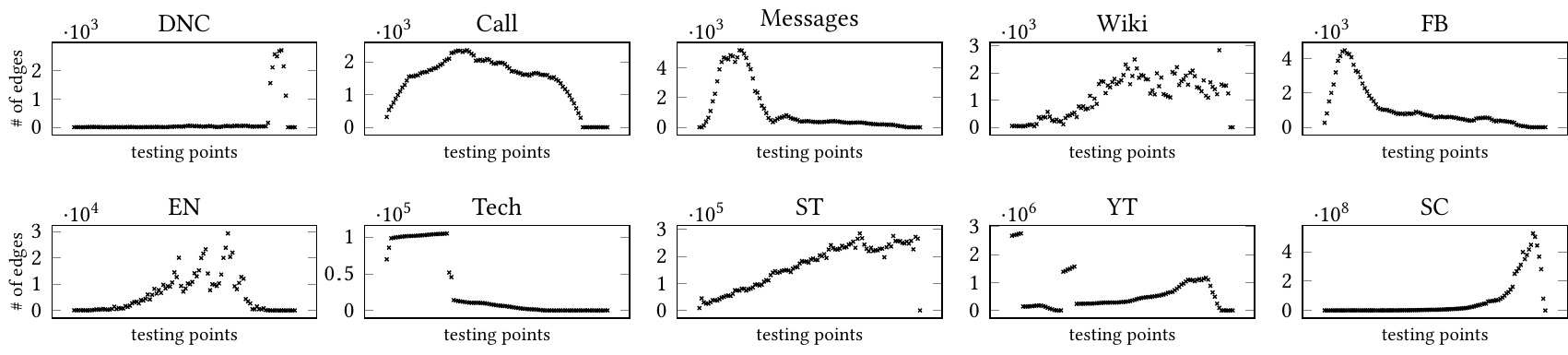}
  \caption{\# edges in graphs.}
  \label{fig:num_edges}
\end{figure*}

\subsection{Average distance between all nodes and roots in spanning trees}
The average distance between all nodes and the 
root in the spanning tree is equal to $\frac{S_d}{|V|}$ where $|V|$ is 
the number of nodes in the spanning tree. Figure~\ref{fig:average_dist} shows the
average distance between all nodes and roots in the spanning trees for
large graphs at each snapshot. Such average distances in D-trees 
are the smallest (all less than 10) and the most stable at each testing 
point in all large graphs. 

\begin{figure*}[htb] \centering
  \begin{subfigure}{\textwidth}\centering
    \includegraphics{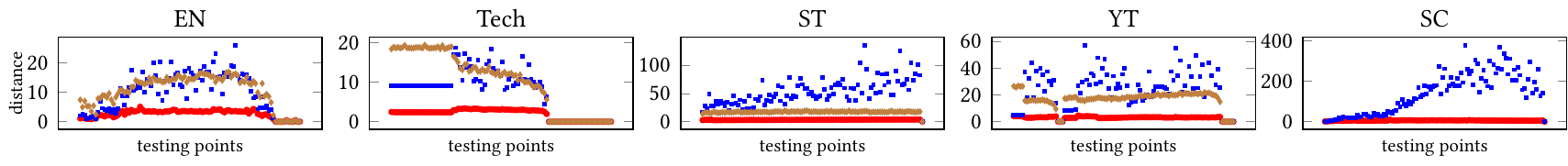}
  \end{subfigure}
  \caption{Average distance between nodes and roots in spanning trees (forests) for large graphs}
  \label{fig:average_dist}
\end{figure*}

\subsection{Average cut number in spanning trees (forests) 
for graphs at each snapshot}

The average cut number in the spanning tree is equal 
to $\frac{S_c}{|V|}$ where $|V|$ is 
the number of nodes in the spanning tree. 
Figure~\ref{fig:average_cutnumber} shows the
average cut numbers in the spanning trees for large graphs at each snapshot. 
Average cut numbers in D-trees are the smallest (all less than 15, in most cases less than 10) 
at each snapshot in all large graphs. 

\begin{figure*}[htb] \centering
  \includegraphics{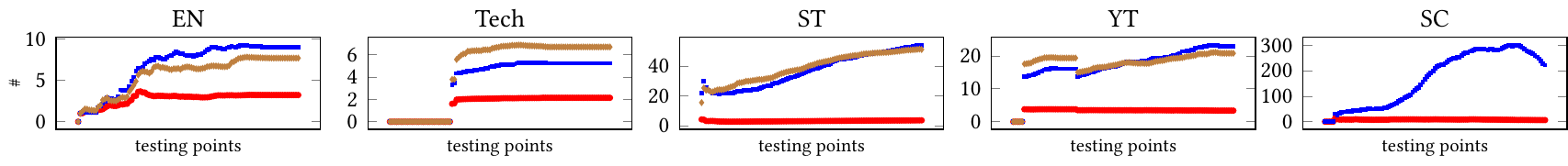}
  \caption{Average cut numbers for spanning trees (forests) for large graphs}
  \label{fig:average_cutnumber}
\end{figure*}

\subsection{Performances of update operations at each snapshot}
\label{updates_performance_at_each_testing_point}
Figure~\ref{fig:updates_over_testing_points} shows performances of update operations
between current snapshot and previous snapshot. Updates on larger graphs take more 
time than on smaller graphs. Overall, D-tree has the best update performances.

\begin{figure*}[htb] \centering
  \begin{subfigure}{\textwidth}\centering
      \includegraphics{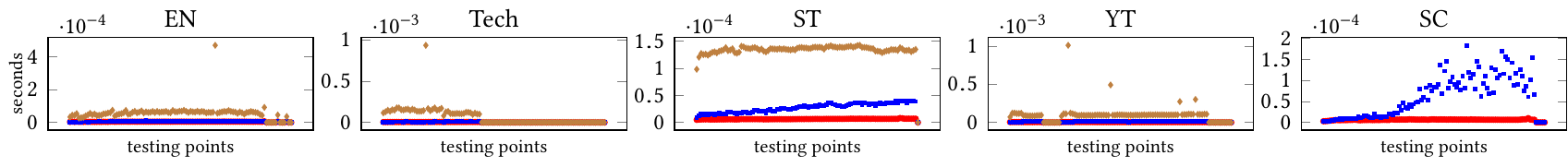}
      \caption{Average run time of inserting tree edges}
  \label{fig:in_te}
  \end{subfigure}
  \begin{subfigure}{\textwidth}\centering
    \includegraphics{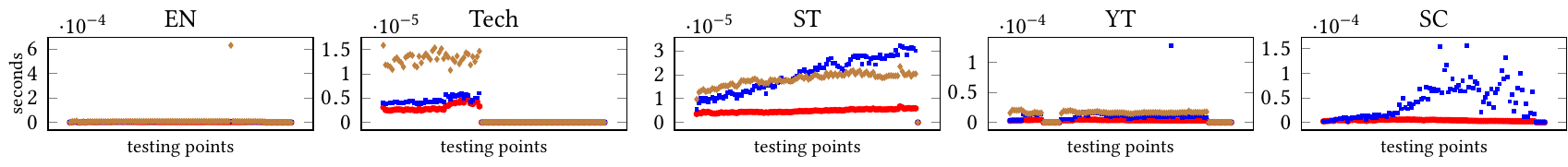}
    \caption{Average run time of inserting non-tree edges}
  \label{fig:in_nte}
  \end{subfigure}
  \begin{subfigure}{\textwidth}\centering
    \includegraphics{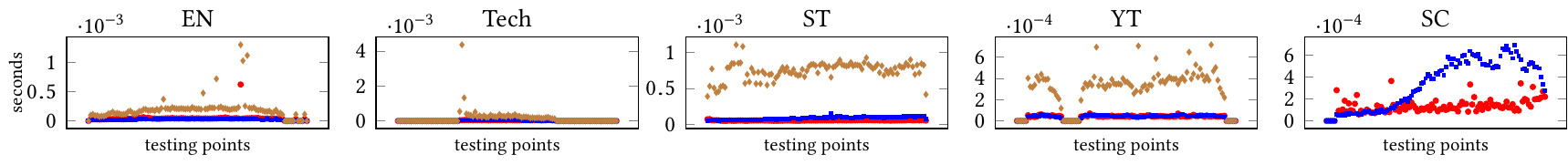}
    \caption{Average run time of deleting tree edges}
  \label{fig:de_te}
  \end{subfigure}
  \begin{subfigure}{\textwidth}\centering
    \includegraphics{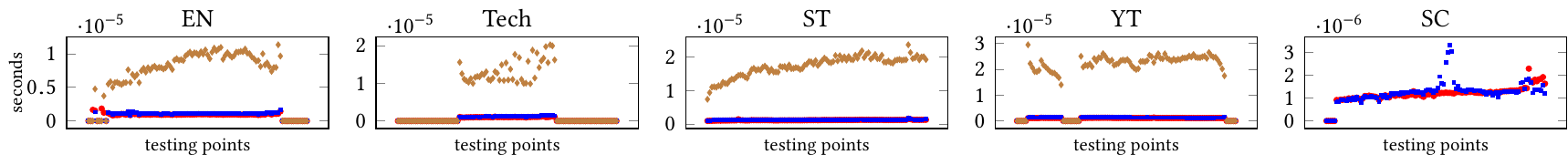}
    \caption{Average run time of deleting non-tree edges}
  \label{fig:de_te}
  \end{subfigure}
\caption{Average run time of update operations between current snapshot
and previous snapshot in forest trees (forests) for large graphs. }
\label{fig:updates_over_testing_points}
\end{figure*}

}

\end{document}